\documentclass{aa}  
\usepackage{natbib}
\bibpunct{(}{)}{;}{a}{}{,}
\usepackage{graphicx}
\usepackage{txfonts}
\usepackage{xcolor, soul}
\sethlcolor{green}

\usepackage{csquotes}
\newcommand{\teff}  {T$_\mathrm{eff}$}
\newcommand{\logg}  {$\log \textit{g}$}
\newcommand{\loggf} {$\log \textit{gf}$}
\newcommand{\kms}   {\mbox{${\rm km}\:{\rm s}^{-1}$}}
\newcommand{\sfe}   {$[\mathrm{S/Fe}]$}
\newcommand{\feh}   {$[\mathrm{Fe/H}]$}
\newcommand{\vsini} {\textit{v}sin\textit{i}}
\newcommand{\vmac}  {$v_\mathrm{mac}$}
\newcommand{\vmic}  {$v_\mathrm{mic}$}
\newcommand{\hamr}  {h$\alpha$mr}
\newcommand{\alphafe}   {[$\alpha$/Fe]}

\begin{document} 

   \title{Chemical abundances of 1111 FGK stars from the HARPS-GTO planet search sample}

   \subtitle{III. Sulfur\thanks{Based on observations collected at the La Silla Observatory, ESO (Chile), with the HARPS spectrograph at the 3.6 m ESO telescope (ESO runs ID 72.C-0488, 082.C-0212 and 085C.0063).}}

   \author{A. R. Costa Silva \inst{1,2}
          \and E. Delgado Mena \inst{1}
          \and M. Tsantaki \inst{1}
          }

   \institute{Instituto de Astrofísica e Ciências do Espaço, Universidade do Porto, CAUP, Rua das Estrelas, 4150-762 Porto, Portugal\\
              \email{ana.rita@astro.up.pt}
              % can I use this email ???
         \and
             University of Hertfordshire, School of Physics, Astronomy and Mathematics, College Lane Campus, Hatfield, Hertfordshire,
             \mbox{AL10 9AB}, UK
             }

   \date{Received ...; accepted ...}

% \abstract{}{}{}{}{} 
% 5 {} token are mandatory
 
  \abstract
  % context heading (optional), leave it empty if necessary  
   {Elemental abundances are of prime importance to help us reconstruct the origin and evolution of stars and galaxies in our Universe. Sulfur abundances have not been as heavily studied as other elements, so some details regarding its behaviour are still unclear.}
   {We aim to investigate \sfe\ ratios in stars of the solar neighbourhood in order to analyse the chemical evolution of sulfur and probe for possible differences in abundances of planet host and non-planet host stars.}
   {We use the code MOOG to perform spectral synthesis and derive \vsini\ values and \sfe\ ratios for 719 FGK stars with high-resolution (R $\sim$115000) and high-quality spectra from the HARPS-GTO program. We find the best fit and corresponding parameter values by performing $\chi^2$ minimisation of the deviation between synthetic profiles and observational spectra.}
   {Our results reveal that sulfur behaves as a typical $\alpha$-element, with low abundances in young thin disk stars and high abundances in old thick disk stars, following what was expected from our understanding of the Galactic chemical evolution (GCE). Nevertheless, further studies into the abundances of sulfur in very metal-poor stars are required as our sample only derived sulfur abundances to stars with metallicity as low as \feh\ $=-1.13$ dex. High-$\alpha$ metal rich stars are more enhanced in sulfur compared to their thin disk counterparts at the same metallicity. We compare our results to GCE models from other authors in the \sfe\ vs. \feh\ plane. The \sfe-age relationship is a good proxy for time, just like it is the case with other $\alpha$-elements. We report no differences in the abundances of sulfur between stars with and without planetary companions in the metallicity range \feh\ $\geqslant -0.3$ dex.}
   {}

 % not absolutely necessary though

   \maketitle
%
%%%%%%%%%%%%%%%%%%%%%%%%%%%%%%%%%%%%%%%%%%%%%%%%%%%%%%%%%%%%%%%%%%%%%%

\section{Introduction}
The early Universe was $\alpha$-enriched because type II supernovae (SNe II) occurred on a much faster time-scale than type Ia supernovae (SNe Ia), and whereas the former produce a great amount of $\alpha$-elements and less iron, the latter are the main production sites of Fe-peak elements \citep[e.g.][]{edvardsson1993}. And so iron-group elements found their way to the interstellar medium (ISM) much later than those formed by $\alpha$-captures. With this in mind, the ratio of $\alpha$-elements over iron ([$\alpha$/Fe]) against metallicity can be interpreted as a cosmic clock depicting some aspects of the Galactic chemical evolution \citep[GCE,][]{haywood2013, buder2018, dmena2019}.

Sulfur is an $\alpha$-element produced in SNe II by hydrostatic and explosive burning of oxygen and silicon \citep{woosleyweaver1995, fenner2004}. For long, it was a neglected element in the study of GCE due to the large uncertainties associated with its derivation, and so other $\alpha$-elements were preferred, such as Si and Ca, which have similar nucleosynthesis. Be that as it may, sulfur is a volatile element, and as such, does not form dust grains in the ISM \citep{savagesembach1996}, making it an ideal depletion-independent tracer of the evolution of galaxies.

In the early 2000s, different trends were reported with regard to the evolution of sulfur. They all agreed that \sfe\ was solar around \mbox{\feh\ $=0$} dex and increased with decreasing metallicity, down to \mbox{\feh\ $=-1.0$} dex. However, in the metal-poor regime, the findings did not reach a consensus. Earlier studies found \sfe\ ratios rising up to $\sim$0.8 dex \citep{israelian2001, takada-hidai2002} or a bimodal behaviour with both high ratios and ratios on a plateau at $\sim$0.3 dex \citep{caffau2005}, but a majority of authors proposed an $\alpha$-like behaviour, showing only ratios on a plateau around 0.3 dex \citep[e.g.][]{chen2002, ryde2004}. This trend has been confirmed by most recent studies \citep[e.g.][and references therein]{duffau2017, caffau2019}.

In terms of the elemental abundances of the Milky Way's populations, it has been found that $\alpha$-elements are more abundant in the thick disk rather than the thin disk \cite[e.g.][]{navarro2011, adibekyan2011, bensby2014}. However, analysis into how sulfur abundances are distributed according to Galactic populations are scarce, thus we aim to analyse if sulfur shows a Galactic distribution similar to that of other $\alpha$-elements by using the HARPS-GTO sample.

Additionally, sulfur is also considered an important building block of terrestrial planets, alongside Fe, Mg, Si, C, and O \citep{bond2010}. It is present in the core of rocky planets, usually in the form of iron alloy, e.g. FeS, and possibly in the mantle, although in smaller amounts \citep{brugger2017, santerne2018, wang2019}. Recent studies \citep{dorn2015, santos2015} into the composition and internal structure of terrestrial planets have concluded that host stellar abundances can be used to set additional constraints when modelling the interiors of these planets, helping to reduce the degeneracies of the models. The planet bulk abundance can generally be assumed to be the same as the star for refractory elements, but not for volatile elements, such as sulfur, carbon or oxygen. In these cases, as the planet is devolatilised, the extent of volatile depletion has to be estimated according to distance from the star and the bulk abundances of the planet can then be inferred from the host stellar abundances \citep[see][and references within]{wang2019}. 

This paper is a continuation of the work started by \citet[][hereafter A12]{adibekyan2012a} and followed by \citet[][hereafter DM17]{dmena2017}, where precise chemical abundances were derived for more than 1000 stars in the solar neighbourhood, although we employ spectral synthesis to fit observational spectra instead of the Equivalent Widths technique. It is structured in the following way: in Section 2, we will characterise the sample and the stellar parameters; Section 3 describes the process of deriving sulfur abundances and the associated uncertainties; in Section 4, we present and discuss our results in the context of GCE, compare them to theoretical models and study the differences in sulfur abundances between planet hosting and non-planet hosting stars; finally, Section 5 will summarise our conclusions.

%%%%%%%%%%%%%%%%%%%%%%%%%%%%%%%%%%%%%%%%%%%%%%%%%%%%%%%%%%%%%%%%%%%%%

\section{Sample description and stellar parameters}

The original sample was composed of 1111 FGK stars observed in the context of three HARPS-GTO planet search programs \citep{mayor2003, locurto2010, santos2011}. The majority of the stars are slow rotators, non-evolved, and have low chromospheric activity levels. The individual spectra of each star were reduced using the HARPS pipeline and then combined with IRAF\footnote{IRAF is distributed by National Optical Astronomy Observatories, operated by the Association of Universities for Research in Astronomy, Inc., under contract with the National Science Foundation, USA.} after correcting for its radial velocity shift. The resolution is $\sim$115 000 and the signal-to-noise ratios (S/N) range from $\sim$20 to $\sim$2000, although most of the sample has high S/N (85\% of the sample has S/N higher than 100).

Stellar parameters were derived in a homogeneous way for the entire sample in \citet{sousa2008, sousa2011a, sousa2011b}. Later, the parameters for cooler stars (\teff $<$ 5200 K) were re-determined \citep[][DM17]{tsantaki2013} using a reduced list of iron lines specially assembled to avoid blending effects. DM17 also corrected the spectroscopic gravities (\logg) of the sample. Abundance references were derived from a solar reflected light spectrum of the Vesta asteroid, also obtained with the HARPS spectrograph. For further details on the sample and parameter determination, see A12 and DM17. 

The final result was a sample of 1059 solar neighbourhood stars, with effective temperatures from 4393 K to 7212 K (98\% of the stars have \mbox{4500 K $<$ \teff $<$ 6500 K}), surface gravities from 2.75 to 5.06 dex (96\% stars have \mbox{4 $<$ \logg $<$ 5 dex}), and metallicities ranging from -1.39 to 0.55 dex. This sample contains 150 planet hosts (29 Neptunian-mass planets or Super-Earths and 121 Jupiter-like planets) and 909 stars which are not known to harbour any planetary companions\footnote{Due to the HARPS radial velocity precision and the long-term duration of the survey, we can only be sure that these stars do not host giant planets. However, we cannot guarantee that they are not hosts to smaller planets, specially those with long periods.}.

Chemical abundances of other elements have been derived in an homogeneous way for this sample by other authors: refractory elements \mbox{\textit{A}$<$29} (A12), lithium \citep{dmena2014, dmena2015}, carbon \citep{sandres2017}, oxygen \citep{bdlis2015}, and neutron capture elements (DM17). Additionally, for a fraction of the stars in this sample, beryllium abundances can be found in \citet{santos2004a, galvez-ortiz2011, dmena2011, dmena2012} and nitrogen abundances in \citet{sandres2016}.

%%%%%%%%%%%%%%%%%%%%%%%%%%%%%%%%%%%%%%%%%%%%%%%%%%%%%%%%%%%%%%%%%%%%%

\section{Derivation of chemical abundances}
Synthetic spectra were created assuming local thermodynamic equilibrium (LTE) using the 2017 version of the spectral synthesis code MOOG \citep{sneden1973}, with the driver \textit{synth}. The stellar atmospheres models are from \citet{kurucz1993}, being the same ones used to derive stellar parameters and other elements' abundances. The line lists for the spectral regions were taken from VALD \citep{kupka-ryabchikova1999}, although the atomic parameters of some lines had to be slightly changed to better match the solar spectra (see more in Section 3.2).

\subsection{Rotational velocity and macroturbulence}
The values of rotational projected velocity (\vsini) and macroturbulence (\vmac) of the stars are inputs required by the MOOG code as they take part in the broadening of the atomic lines, so in order to derive sulfur abundances, we first had to compute these two parameters.

Macroturbulence values for stars in the temperature range of 5000K to 6500K were determined with Equation (8) from \citet{doyle2014}, which is a temperature and gravity dependent formula. Stars with temperature below 5000K were assigned a value of 2.0\kms\ and for those above 6500K, \vmac\ was set to 5.5\kms. 

As for the rotational velocity, we produced synthetic spectra with MOOG (using the previously derived \vmac\ value) and fitted it to the observed spectra in 36 wavelength regions where over 70 FeI and FeII lines are present. We used the Python package \mbox{\textit{mpfit}} \citep{markwardt2009} to find the best fit and corresponding \vsini\ value by fixing the \feh\ abundances and all other stellar parameters and setting \vsini\ as the only free parameter. The package performs $\chi^2$ minimisation of the deviation between the synthetic and observed data sets in order to find the best-fitting values of the free parameter defined by the user. Errors are automatically computed by the package as well.  

For ten stars in our sample (HD82342, HD55, HD108564, HD145417, HD134440, HD23249, HD40105, HD92588, HD31128, HD52449), the value of \vmac\ that was first attributed to them caused a malfunction in the code that determined \vsini, possibly due to \vmac\ being too large a value, as it tried to compensate by finding \vsini\ values that were below 0.1 \kms. Therefore the \vmac\ of these stars were empirically adjusted for the code to work correctly. For some stars, the minimisation did not converge.

As a way of testing it, we computed values for the Sun (\teff=5777K, \logg=4.39, \feh=0.02) using the Vesta-reflected spectrum, for which we got \mbox{\vmac = 3.31 \kms} and \mbox{\vsini = 2.0$\pm$0.1\kms}. In \citet{doyle2014}, the \vsini\ adopted for the Sun was 1.9 \kms, which is in close agreement with the value computed in our study.

% *******************
\subsection{Sulfur abundances}

% -------------------------- FIGURE ------------------------------
\begin{figure*}
    \centering
    \includegraphics[width=0.95\hsize]{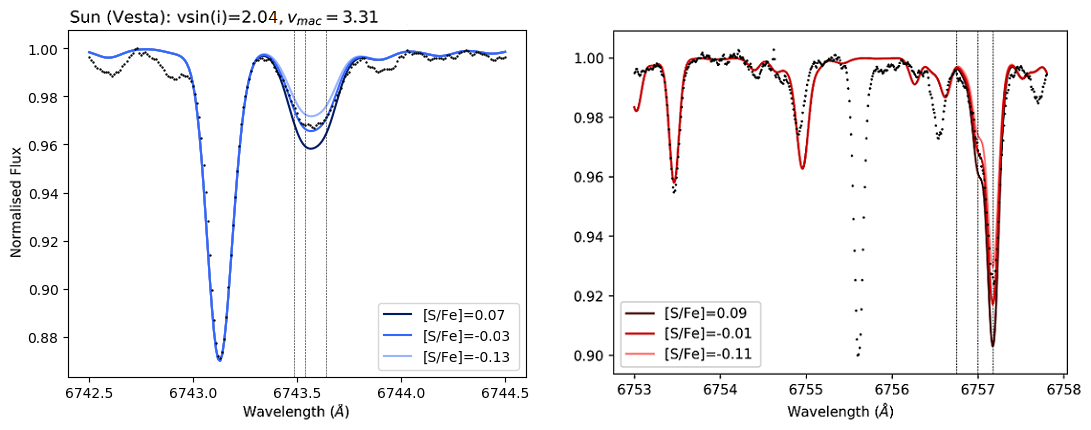}
    \includegraphics[width=0.95\hsize]{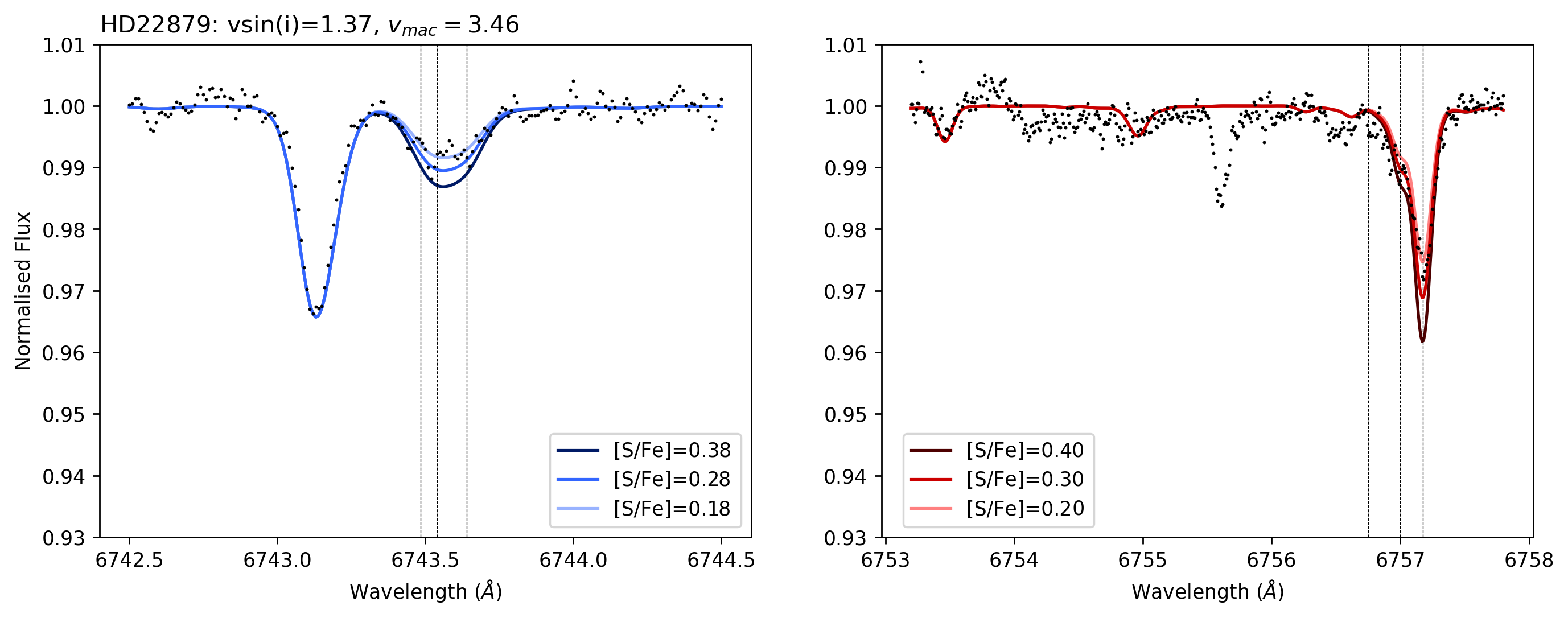}
    \caption{Example of fitting profiles around \mbox{6743 \AA} (left) and \mbox{6757 \AA} (right). \textit{Top panels:} solar reflected spectrum from Vesta. \textit{Bottom panels:} spectrum of the metal-poor star HD22879 (\teff=5884K, \logg=4.45 dex, \feh $=-0.82$ dex, \vmac=3.46\kms , \vsini=1.37\kms). The synthetic profiles are for the best [S/Fe] value determined (before correction with respect to solar value) and for $\pm0.1$ dex deviation from it.}
    \label{fits}
\end{figure*}
% --------------------------------------------------------------

% ---------------------------- TABLE ---------------------------
\begin{table}
    \caption{Atomic parameters for the S lines used in this work (left to right): wavelength, excitation energy, oscillator strength, Van der Waals damping constant, literature reference for data. All lines belong to Multiplet 8.}
    \label{table:atomic:parameters}
    \centering
    \begin{tabular}{c c c c c}
        \hline\hline
        $\lambda$ & $\chi_{l}$ & \loggf & log($\gamma_{6}/\mathrm{N}_{H}$)  & Reference \\
        (\AA) & (eV)  & &  ($\mathrm{s}^{-1}\mathrm{cm}^3$) & \\
        \hline
        6743.4834  &   7.866  &  -1.27  &  -7.16 & \textit{emp.} \\
        6743.5400  &   7.866  &  -0.95  &  -7.16 &  \cite{kurucz2004}\\
        % 6743.5801  &   7.866  &  -4.85  &  -7.16 \\
        % 6743.5801  &   7.866  &  -4.20  &  -7.16 \\   
        6743.6401  &   7.866  &  -0.93  &  -7.16 & \textit{emp.}\\
        6756.7500  &   7.870  &  -1.67  &  -7.16 & \cite{kurucz2004}\\
        % 6756.8511  &   7.870  &  -1.76  &  -7.16 \\
        6757.0000  &   7.870  &  -0.83  &  -7.16 & \cite{kurucz2004}\\
        6757.1750  &   7.870  &  -0.24  &  -7.16 & \cite{kurucz2004}\\
        \hline
    \end{tabular}
\end{table}
%-----------------------------------------------------------
%
The determination of sulfur abundances was performed with the same package \textit{mpfit}, now setting [S/Fe] as the only free parameter. We fitted synthetic profiles obtained with MOOG to observational spectra of two \ion{S}{i} atomic lines from Multiplet 8\footnote{Numbering as in \citet{moore1945}.}, around \mbox{6743 \AA} and \mbox{6757 \AA}. Atomic parameters can be found in Table \ref{table:atomic:parameters}, and were obtained from \citet{kurucz2004, wiese1969}. The lines at wavelengths 6757.000 \AA\ and 6757.175 \AA\ were originally at 6756.960 \AA\ and 6757.150 \AA, respectively, and for the lines at 6743.483 \AA\ and 6743.640 \AA, the values of \loggf\ were slightly adapted. These changes were carried out so that the lines would better match the Kurucz Atlas solar spectrum and we could obtain A(S)$_{\odot}$=7.12 dex. A third S line around \mbox{6748 \AA} was considered, but inaccuracy in the line list ultimately prevented us from obtaining reliable abundances, and hence we discarded this line.

\citet{korotin2008, korotin2009} has studied departures from LTE and found the effects to be negligible on Mult. 8 (corrections are smaller than -0.1 dex). These lines are formed in denser layers of stellar atmospheres and are therefore less sensitive to non-LTE (NLTE) effects.

The $\chi^2$ analysis was performed separately in the intervals \mbox{6743.3-6743.8 \AA} and \mbox{6756.7-6757.4 \AA}, and the final [S/Fe] value is the mean of the two individual values, after being independently corrected with respect to the solar value.

Upon visual inspection of the fitting profiles, we decided to establish a cut-off temperature at \mbox{5000 K} because the spectra of cool stars were showing multiple lines that we could not reproduce due to incompleteness of the line list. Moreover, the used S lines have a high excitation potential and become very weak as the star becomes cooler. Unfortunately, these problems could not be solved and these stars had to be ruled out from the sample. We kept ten stars with temperatures below the cut-off \teff\ (although all with \teff$>$\mbox{4900 K}) whose synthetic profile was deemed to be accurate. Furthermore, we applied multiple S/N cut-offs for different \teff\ ranges to make sure our final results are as reliable as possible.

We tested our code by deriving the sulfur abundances of the Sun and found [S/Fe]$_{average}=\mathrm{-0.02}\pm\mathrm{0.02 dex}$. Figure \ref{fits} exemplifies the fitting of the synthetic profiles to observational data in both wavelength regions, for the solar spectrum and for the spectrum of the metal-poor star HD22879 (\feh $=-0.82$dex). After the fittings, all of the sulfur ratios derived for our sample were corrected with respect to the solar value obtained in the corresponding S line, and the average value was calculated afterwards.

% **********************
\subsection{Uncertainties}

% --------------------------- FIGURE -----------------------
\begin{figure}
    \centering
    \includegraphics[width=0.95\hsize]{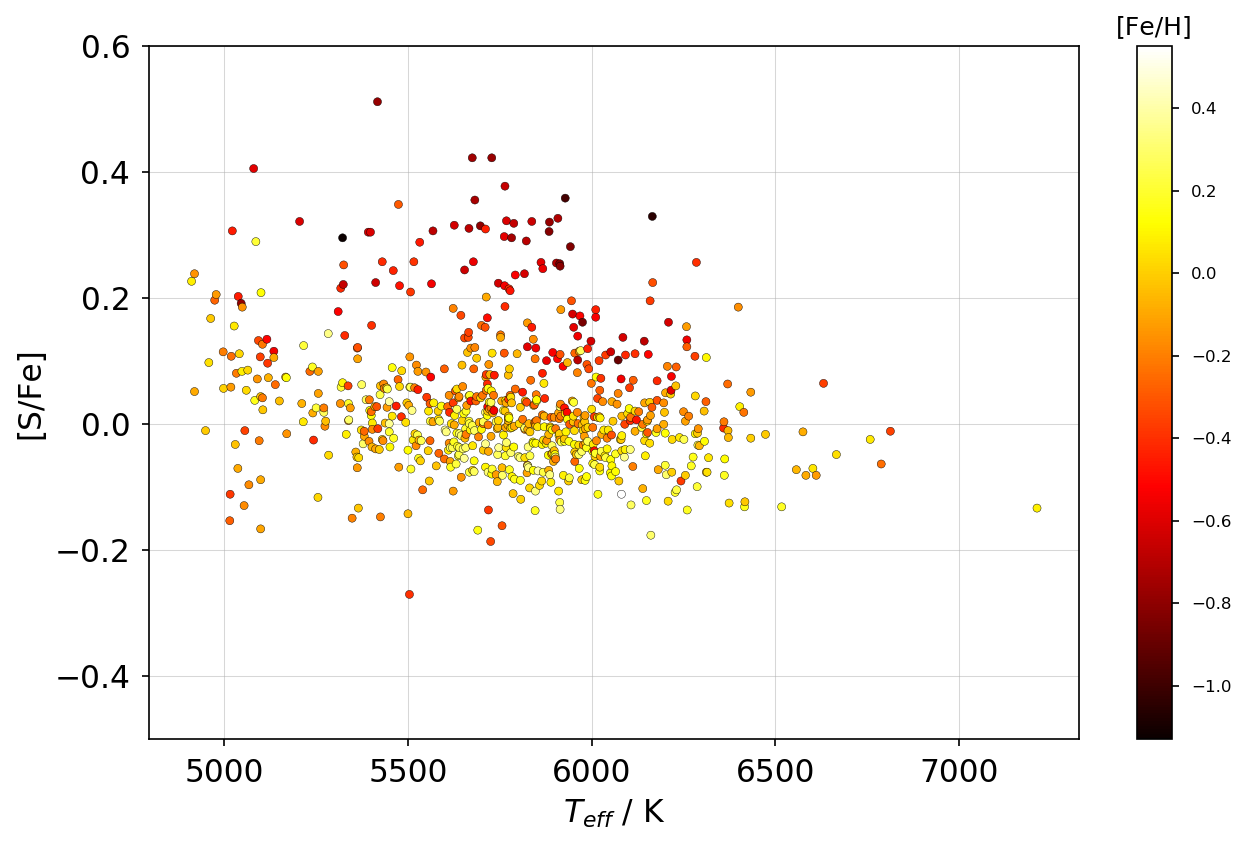}
    \includegraphics[width=0.95\hsize]{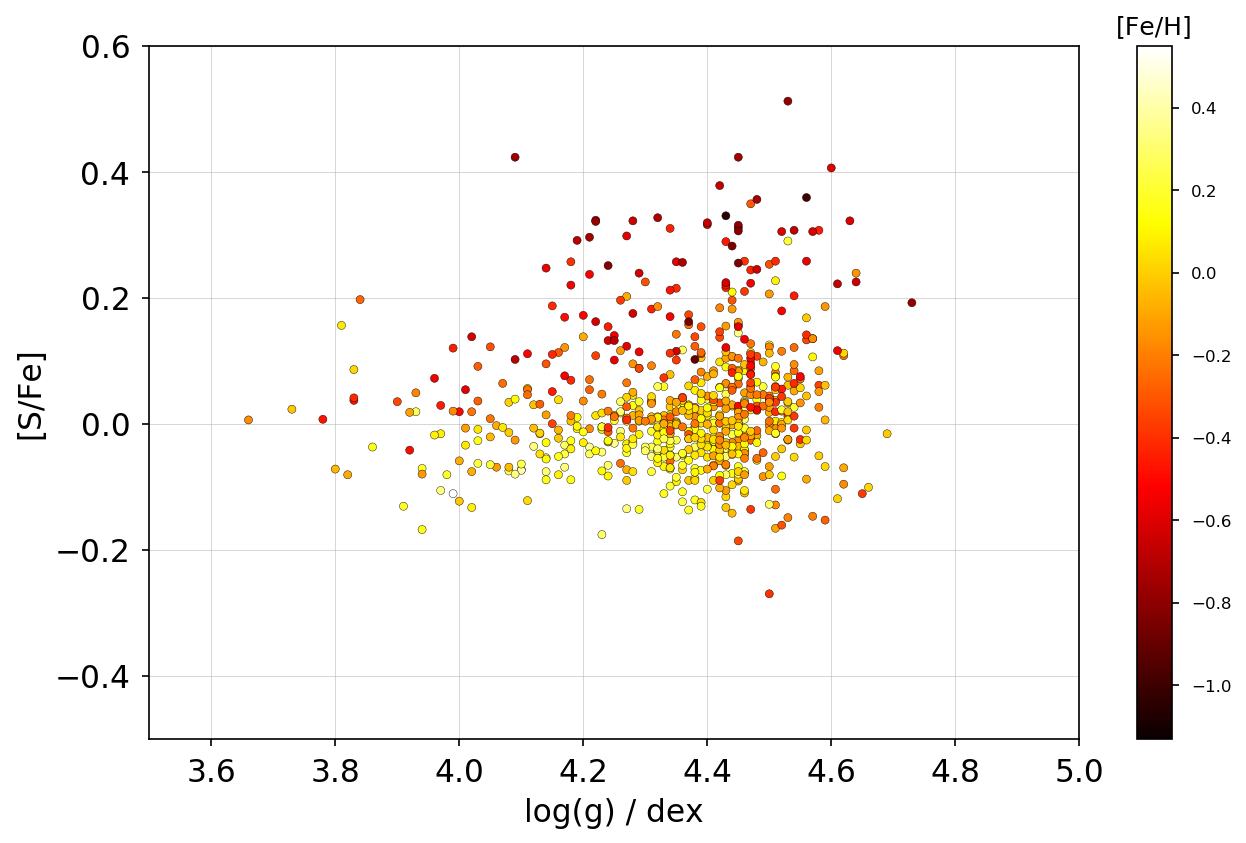}
    \caption{Relation of \sfe\ with \teff\ (top) and \logg\ (bottom), colour
    coded by metallicity. No obvious trends are detected.}
    \label{parameter:trends}
\end{figure}
% ---------------------------------------------------------

% ---------------------------- TABLE ---------------------
\begin{table*}
    \caption{Sensitivity of the [S/Fe] ratio (in dex) due to errors on stellar parameter. The values presented are the means within each of the defined temperature subsamples.}
    \label{table:parameter:sensitivity}
    \centering
    \begin{tabular}{l c | l c | l c}
    \hline \hline
    Low \teff & $\Delta$\sfe & Solar \teff & $\Delta$\sfe & High \teff & $\Delta$\sfe \\
    \hline
    $\Delta$\teff$=\pm64$ K & $\pm0.03$ & $\Delta$\teff$=\pm24$ K & $\pm0.01$ & $\Delta$\teff$=\pm46$ K & $\pm0.02$ \\
    $\Delta$\logg$=\pm0.17$ dex & $\pm0.05$ & $\Delta$\logg$=\pm0.03$ dex & $\pm0.01$ & $\Delta$\logg$=\pm0.05$ dex & $\pm0.01$ \\
    $\Delta$\feh$=\pm0.04$ dex & $\pm0.02$ & $\Delta$\feh$=\pm0.02$ dex & $\pm0.02$ & $\Delta$\feh$=\pm0.03$ dex & $\pm0.03$ \\
    $\Delta$\vmic$=\pm0.33$ \kms & $<0.01$ & $\Delta$\vmic$=\pm0.04$ \kms & $<0.01$ & $\Delta$\vmic$=\pm0.08$ \kms & $<0.01$ \\
    \hline
    \end{tabular}
\end{table*}
% ---------------------------------------------------------
%
In this work, the errors in sulfur abundances are the quadratic sum of the line-to-line scatter and the effects that propagated from the errors in the stellar parameters. We note that some uncertainty may also arise from the automatic continuum placing, hence the importance of visually checking the fits and manually correcting, if necessary.

Table \ref{table:parameter:sensitivity} presents the mean errors of the parameters and the mean variation of sulfur when one of the stellar parameters is varied by their individual error. The sample was divided into three subgroups, as was done in A12 and DM17, to account for the dependence of all the parameters on the temperature of the stars, and so stars with \mbox{\teff $<$ 5277 K} are part of the "low \teff" subsample, stars with \mbox{\teff = $T_{\odot}\pm$500 K} belong to the "solar \teff" group, and lastly, "high \teff" includes the stars with \mbox{\teff $>$ 6277 K}.

As expected, the "low \teff" group is the most sensitive to errors in the parameters, whilst the "solar \teff" is the least. We can also see that the errors in microturbulence are practically negligible for the chosen lines.

%%%%%%%%%%%%%%%%%%%%%%%%%%%%%%%%%%%%%%%%%%%%%%%%%%%%%%%%%%%%%%%%%%%%%%

\section{Results and discussion}

% ---------------------------- TABLE ---------------------
\begin{table*}
    \caption{Sample table of the derived \vmac, \vsini, sulfur abundance, error, number of S lines that were fitted, and signal-to-noise ratio of the spectrum.}
    \label{sample:table:parameters}
    \centering
    \begin{tabular}{l c c c c c c c c c}
    \hline \hline
    Star & \teff\ & \logg\ & \feh\ & \vmac\ & \vsini\ & \sfe\ & $\sigma_{\mathrm{[S/Fe]}}$ & \textit{n}$_{\mathrm{[S/Fe]}}$ & S/N \\
    \hline
    ...   &   ...   &   ...  &  ...  &  ...  &  ...  &  ...  & ...  &  ... & ... \\ 
    HD127124   &   5030.0   &   4.55  &  -0.03  &  2.37  &  0.51  &  -0.03  &  0.10  &  2 & 218 \\ 
    HD13808    &   5033.0   &   4.51  &  -0.21  &  2.44  &  0.50  &   0.08  &  0.17  &  2 & 1015 \\
    HD191797   &   5037.0   &   4.62  &  -0.09  &  2.22  &  4.10  &  -0.07  &  0.15  &  2 & 321 \\
    HD17970    &   5038.0   &   4.54  &  -0.45  &  2.38  &  0.27  &   0.20  &  0.13  &  2 & 338 \\
    ... & ... & ... & ... & ... & ... & ... & ... & ... & ... \\
    \hline
    \end{tabular}
\end{table*}

We successfully derived [S/Fe] ratios for 719 stars from our sample. For 706 of them, the sulfur abundance presented is the mean of the values obtained from the two lines, whereas for the other 13 stars, the abundances could only be derived from one of the lines. Figure \ref{parameter:trends} presents the relationship between \sfe\ and \teff\ and \logg. It is clear that there are no systematic trends of the abundance ratios caused by the \teff\ or \logg\ parameters, but the colour scheme shows a trend of increasing \sfe\ with decreasing \feh. This trend is expected from the GCE, as metal-poor stars usually have higher \sfe\ ratios. Table \ref{sample:table:parameters} shows a sample of our results as well as some of the stellar parameters derived in previous works.

Below, we analyse our results in some different contexts.

% *********************
\subsection{[S/Fe] vs. [Fe/H]}
% --------------------------- FIGURE -----------------------
\begin{figure}
    \centering
    \includegraphics[width=0.95\hsize]{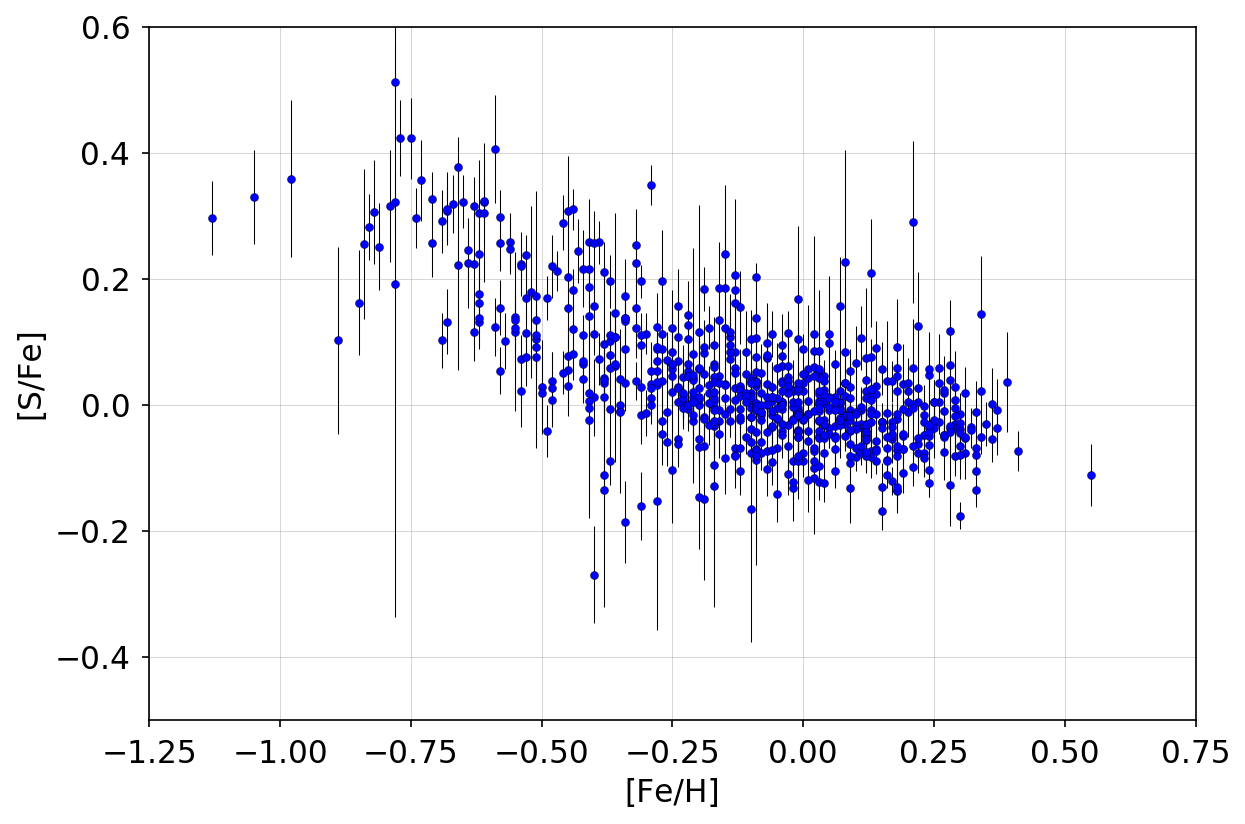}
    \caption{\textit{Top:} \sfe\ vs \feh\ with error bars. The stars with largest errors are usually those that had abundance derived from one line only.}
    \label{sfe_feh}
\end{figure}
% -----------------------------------------------------------

% --------------------------- FIGURE -----------------------
\begin{figure}
    \centering
    \includegraphics[width=\hsize]{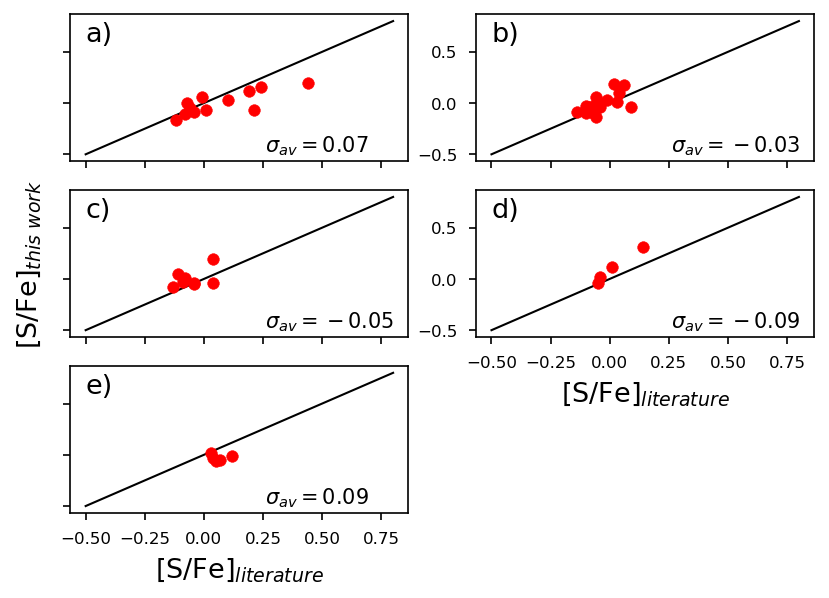}
    \caption{Comparison of [S/Fe] ratios with values found in the literature. Average deviation of all points is displayed in the bottom right. References: a) \cite{maldonado2016}; b) \cite{takeda2016}; c) \cite{mishenina2016}; d) \cite{duffau2017}; e) \cite{caffau2019}.}
    \label{comparison}
\end{figure}
% -----------------------------------------------------------

In Fig. \ref{sfe_feh}, we depict the sulfur abundances ratios against metallicity. The plot includes the total error bars associated with the derivations. The stars with the largest errors are mostly stars that had abundances derived from only one line.

From our results, the abundances of sulfur appear to be decreasing towards higher metallicities, with \sfe\ values tending to $\sim-0.1$ at \feh\ $\sim0.25$. From solar metallicities down to \feh\ $\sim-0.25$, the \sfe\ ratios are close to solar, and with further decreasing metallicity, an increasing trend is very clearly distinguished. At around \feh\ $=-0.75$, despite some scatter, the ratios appear to stabilise and form a plateau at \sfe\ $\sim0.3$. This behaviour is very typical of $\alpha$-elements, and thus has been the theoretical expectation for sulfur.

To help ascertain which of the different trends that have been proposed with respect to the GCE of sulfur is more accurate, we would have to analyse stars below the \mbox{\feh\ $=-1.0$ dex} threshold. Unfortunately, despite the initial sample containing stars with metallicities as low as $-1.39$ dex, the sample for which we could derive \sfe\ ratios only includes two stars with \mbox{\feh\ $\leqslant-1.0$}. This is due to the fact that the lines from Mult. 8 used in this work are difficult to detect in metal-poor stars. Nonetheless, if we look closely at these two metal-poor stars, both of them have \sfe\ values around 0.35, which lends some support to the existence of a flat trend at the metal-poor regime.

In general, our results of \sfe\ vs \feh\ are in good agreement with those of recent studies by \citet[taking into account that NLTE corrections were not applied in this work]{mishenina2015, takeda2016, duffau2017, caffau2019}.

Furthermore, we searched the literature for stars in common with the ones from our sample so as to compare the derived \sfe\ ratios to the ones derived in other works. The comparison of our \sfe\ ratios to those of \cite{maldonado2016, takeda2016, mishenina2016, duffau2017, caffau2019} is plotted in Fig. \ref{comparison}. The samples used by these authors had 12, 19, 8, 4 and 5 stars in common with our own sample, respectively, which yielded an average deviation ($\sigma_{av}$) for each of the comparison works of 0.07, -0.03, -0.05, -0.09 and 0.09 dex, respectively. The values derived in the comparison works are in good agreement with our abundance ratios. Appendix A contains the full table of \sfe\ ratios for each star in common with the aforementioned literature.

% ***************
\subsection{Galactic populations and [S/Fe]-age relation}

% --------------------------- FIGURE ------------------------
\begin{figure}
    \centering
    \includegraphics[width=0.95\hsize]{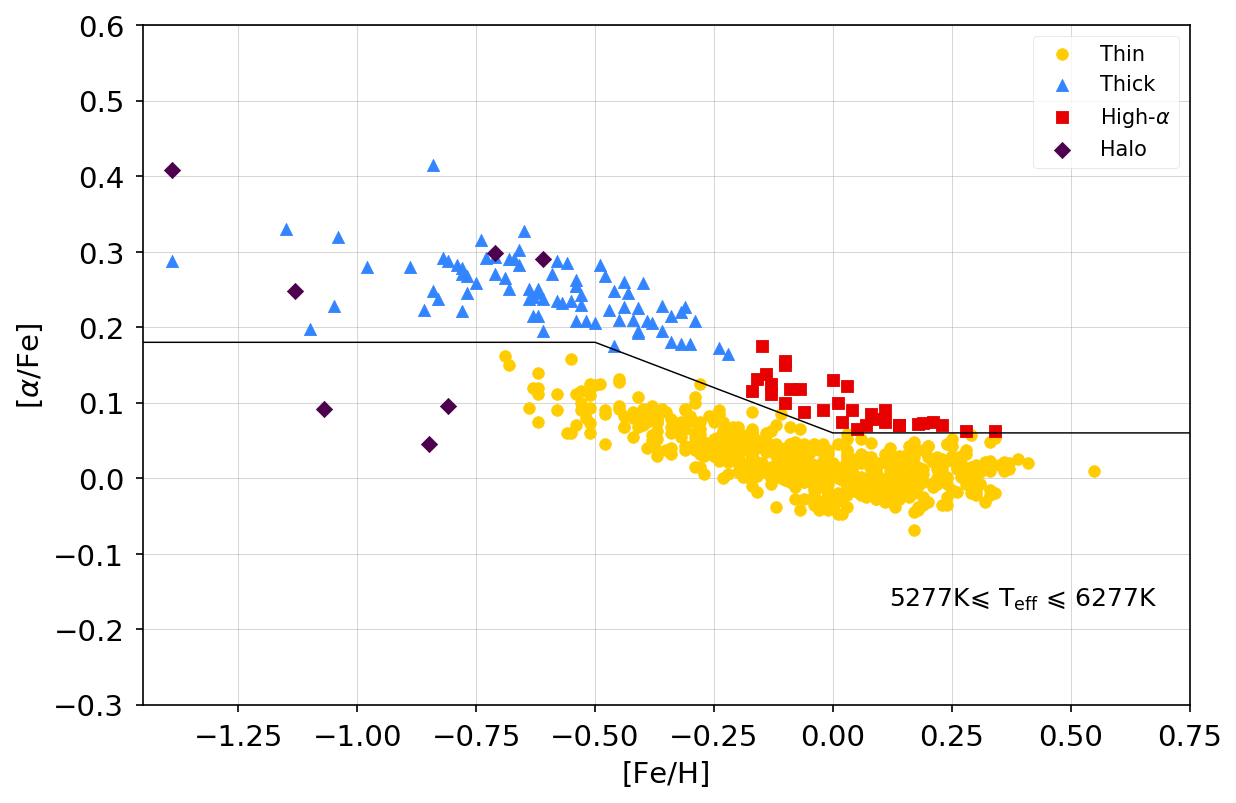}
    \includegraphics[width=0.95\hsize]{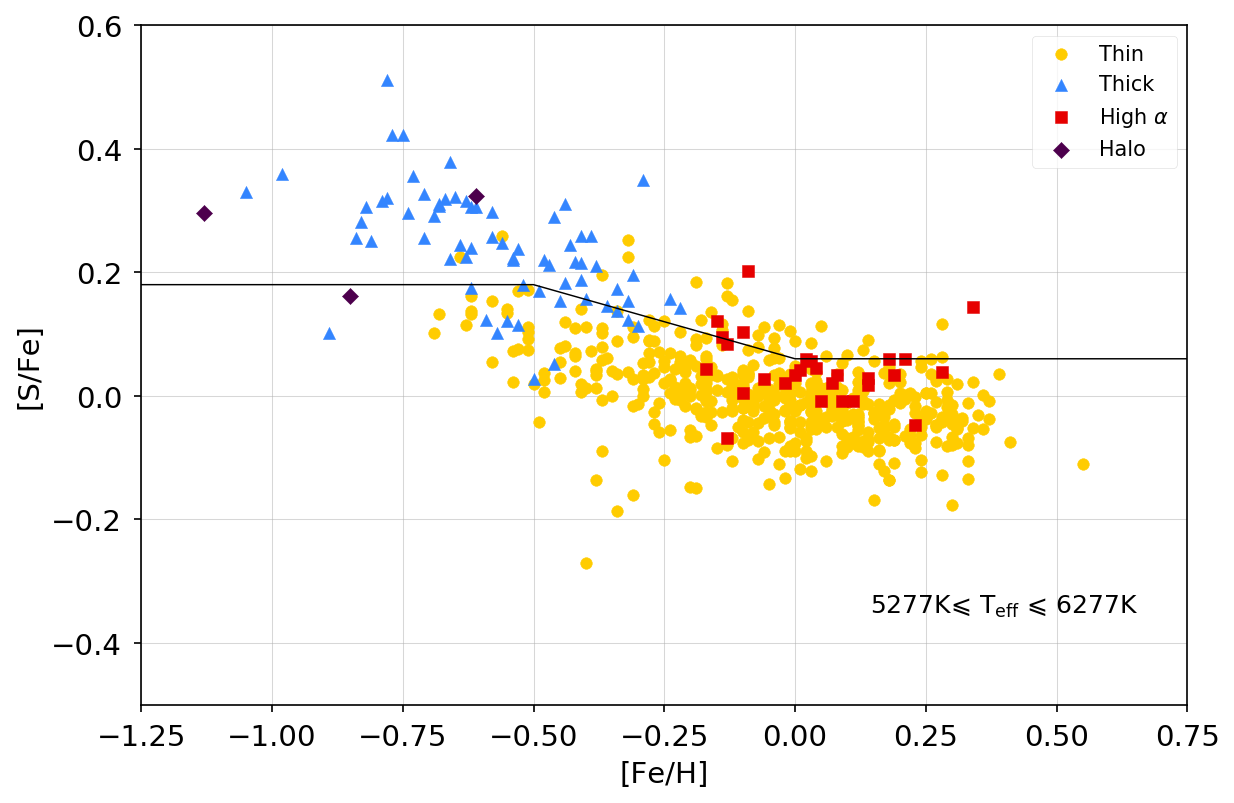}
    \includegraphics[width=0.92\hsize]{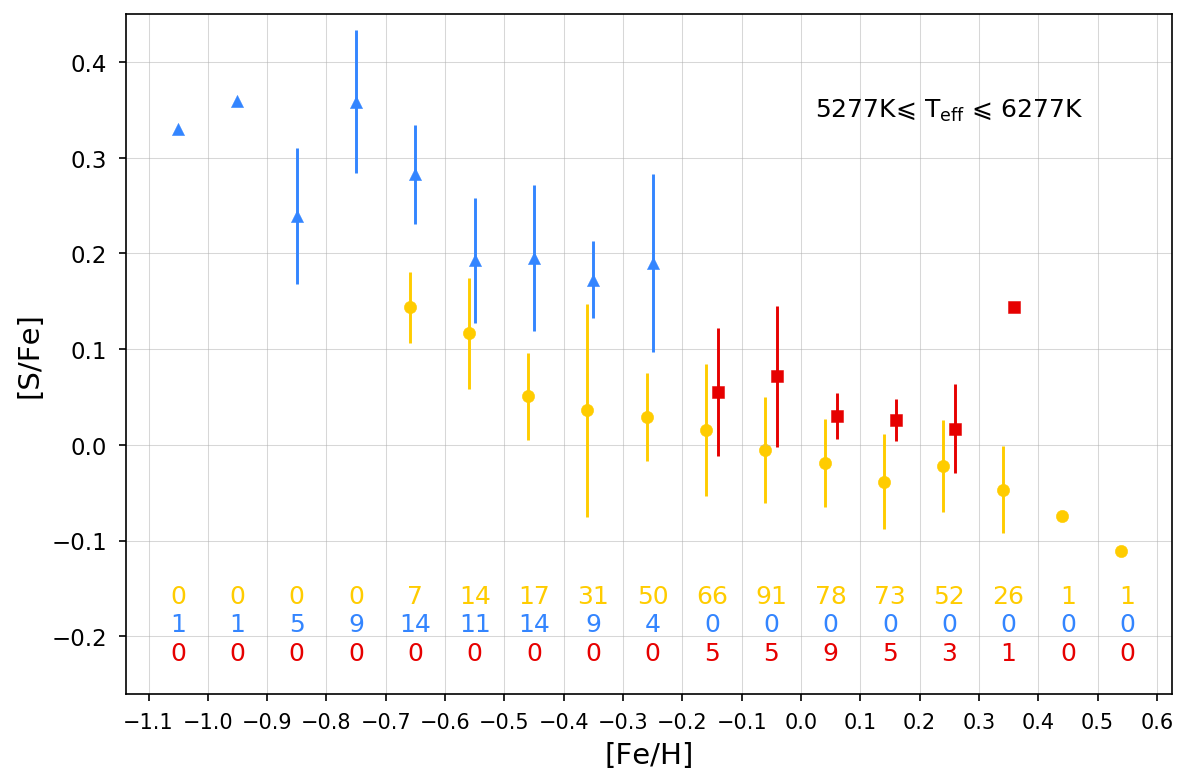} 
    \caption{\textit{Top:} \alphafe\ vs \feh, data from DM17. The black line is the updated $\alpha$ separation line, also from DM17. (see text for more detail). \textit{Middle:} [S/Fe] vs [Fe/H], with the same $\alpha$-line as the top panel. \textit{Bottom:} Average [S/Fe] for each metallicity bin (of width 0.1 dex) per population. The numbers indicate the size of the sample of each population in each bin. \textit{Symbols:} yellow circles are for the thin disk, blue triangles for the thick disk, red squares for the high-$\alpha$ metal-rich stars, and purple diamonds for halo stars.
    }
    \label{sfe:feh:pops}
\end{figure}
% -----------------------------------------------------------

% --------------------------- FIGURE ------------------------
\begin{figure}
    \centering
    \includegraphics[width=0.95\hsize]{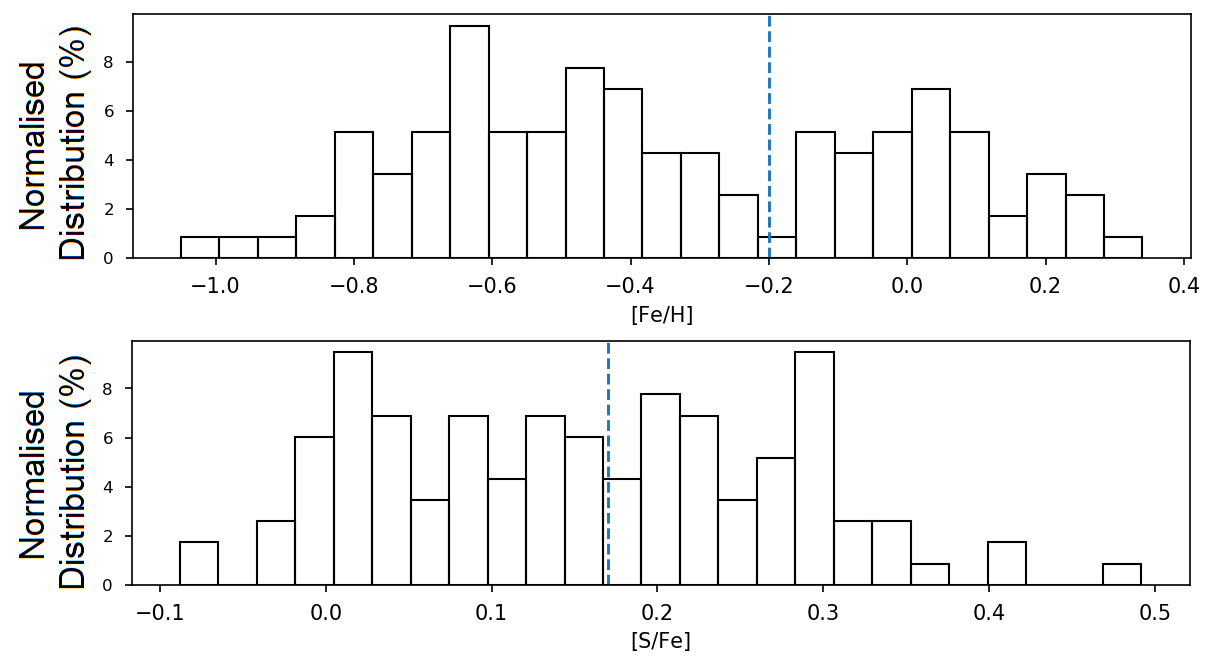}    
    \caption{[Fe/H] and [S/Fe] separation histograms for stars of the thick disk and \hamr\ (high-$\alpha$ sequence). The vertical lines indicate the low-density regions found in \cite{adibekyan2011}, at \mbox{\feh\ $=-0.2$} dex (top histogram) and \mbox{\alphafe\ $\sim0.17$} dex (in the lower histogram, we plot the line at the same value of sulfur abundance: \mbox{\sfe\ $=0.17$} dex).}
    \label{sfe:histograms}
\end{figure}
% -----------------------------------------------------------

% --------------------------- FIGURE ------------------------
\begin{figure}
    \centering
    \includegraphics[width=0.95\hsize]{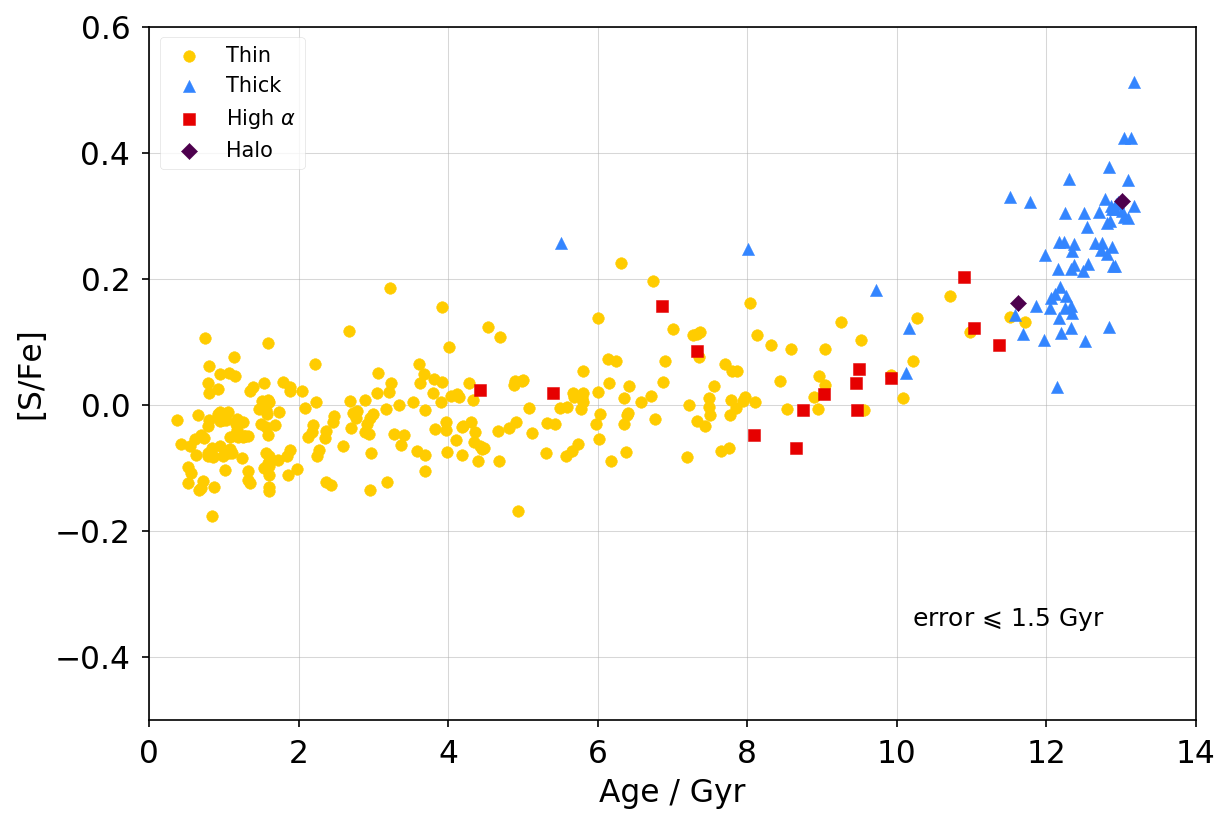}
    \includegraphics[width=0.95\hsize]{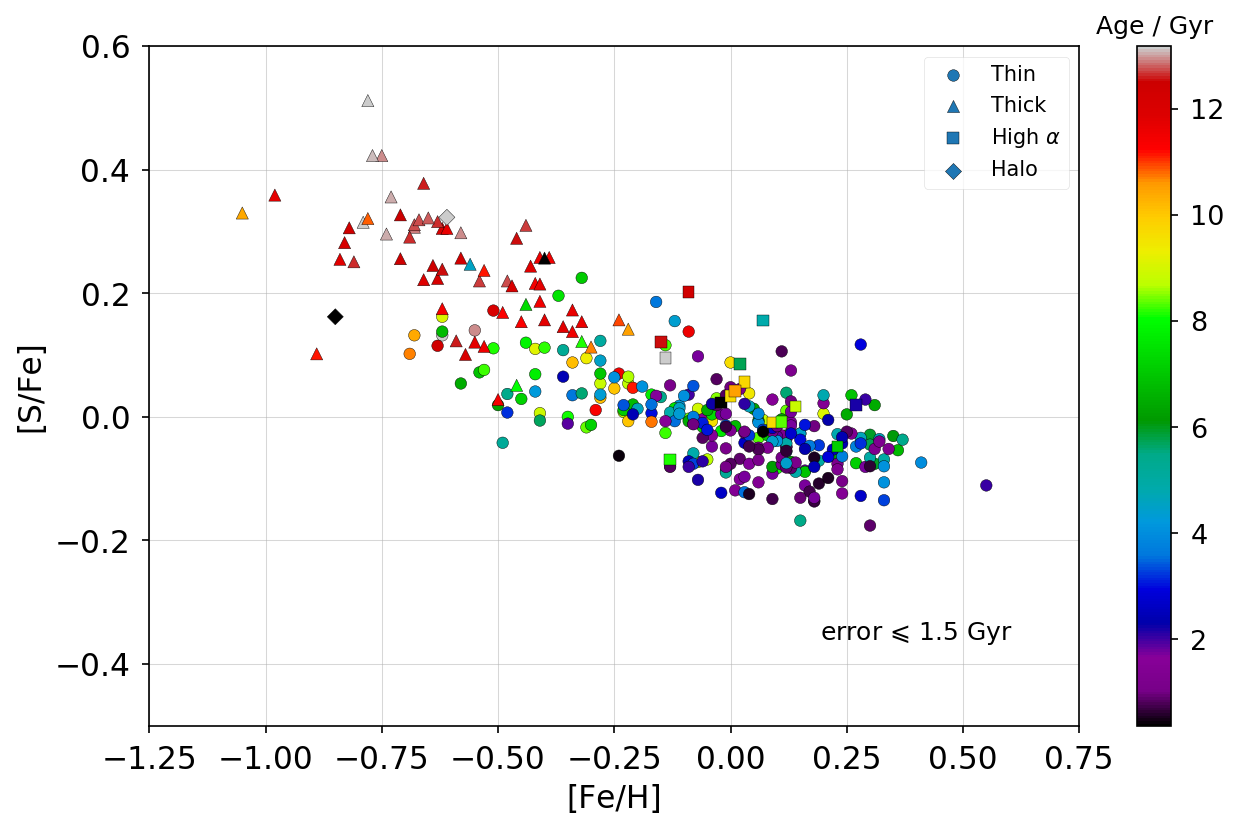}
    \caption{\textit{Top:} [S/Fe] vs age, symbols for galactic populations as in Fig \ref{sfe:feh:pops}. \textit{Bottom:} Distribution of stellar age in the \sfe\ vs \feh\ plane. The sample depicted in both panels is composed only of stars with age uncertainties lower than 1.5 Gyr.
    }
    \label{sfe:age:pops}
\end{figure}
% -----------------------------------------------------------

The Milky Way's thick disk was discovered over three decades ago by \cite{gilmorereid1983}. To the present day, it is still not clear which is the best method to accurately separate stars of the thin and thick disk \citep[e.g. kinematically, chemically, via age, see][and references within]{buder2018}, or even whether more populations (or sub-populations) should be considered \citep{adibekyan2011, chiappini2015, martig2015, rojas-arriagada2016}. 

The stars in our sample were classified as described in \citet{adibekyan2011, adibekyan2013}, according to the chemical separation observed in [$\alpha$/Fe] ($\alpha$ being the average of Mg, Si, and Ti) across different metallicity bins, using the updated abundance ratios derived in DM17. The line that separates the low- and high-$\alpha$ sequences in the \alphafe\ vs \feh\ plane was first presented in \cite{adibekyan2011} and later updated in DM17 (see their Fig. 9). The updated line is shown in Fig. \ref{sfe:feh:pops} and will henceforth be referred to as the $\alpha$-line.
% The regions of low-density in [$\alpha$/Fe], which separate the low- and high-$\alpha$ sequences, were then linked together on the \sfe\ vs. \feh\ plot, creating the line that is shown in Fig. \ref{sfe:feh:pops}, henceforth referred to as the $\alpha$-line. 
Exceptionally, halo stars were categorised on the basis of kinematics. We will refrain from commenting on the abundances of the halo stars as they are only three.

Our sample is composed of 600 thin disk stars, 74 thick disk stars, 42 from the \hamr\ group (explained in the following paragraph), and 3 from the Galactic halo. The different stellar populations are plotted in the middle panel of Fig. \ref{sfe:feh:pops} in the \sfe\ vs \feh\ plane.

Adding to the well-known thin and thick disk populations, the works of \citet{adibekyan2011, adibekyan2013} revealed a potential new group of Galactic disk stars with high metallicity (\mbox{\feh\ $>-0.2$} dex) and enhanced [$\alpha$/Fe] ratios compared to the thin disk, hence called the high-$\alpha$ metal-rich (hereafter \hamr) stars. The \hamr\ and the thick disk populations form the \mbox{high-$\alpha$} sequence, and the separation between the two comes from the analysis of histograms depicting the stellar distributions in [Fe/H] and [$\alpha$/Fe], where well-defined minima (low-density regions) were found (see Fig. \ref{sfe:feh:pops}). More recently, other authors have investigated a possible separation between these two populations, but most often found a continuous distribution rather than clear separations, suggesting that \hamr\ stars were simply the metal-rich tail of the thick disk \citep{recioblanco2014, bensby2014, buder2018}, enriched by both SNe II and SNe Ia. In this paper, we will investigate how the abundances of sulfur fit into the proposed populations.

In the upper panel of Fig. \ref{sfe:feh:pops}, we display the \alphafe\ ratios (where $\alpha$ includes Mg, Si, TiI, and TiII), derived in DM17, against metallicity, for comparison with the \sfe\ ratios from this work, which are depicted in the middle panel, also against metallicity. In both panels, we include only stars with \mbox{\teff = $T_{\odot}\pm$500 K} and we over-plotted the aforementioned $\alpha$-line. The data from DM17 and our own data show similar distributions for each of the populations, although our \sfe\ ratios are more scattered and the populations are less well separated. The $\alpha$-line does not accurately separate the two low- and high-$\alpha$ sequences in the \sfe\ vs \feh\ plane, but it should come as no surprise that not all $\alpha$-elements behave exactly in the same way. In this case, we know sulfur typically has large errors, hence there is a greater deviation and the separation may have become blended as a result.

The bottom panel of the same figure shows the average \sfe\ ratio in each metallicity bin, for each population. As expected, the populations of thick and thin disks clearly follow different $\alpha$-enrichment trends, with thick disk stars dominating the low-metallicity regime with higher \sfe\ ratios, whereas conversely, the thin disk stars are metal-richer but poorer in sulfur. In this panel, we can see that \hamr\ stars do display noticeably higher \sfe\ ratios than the thin disk counterparts at the same metallicity. 

With regard to the separation between thick disk and the \hamr\ group, Fig. \ref{sfe:histograms} shows histograms of the same type as those analysed in \cite{adibekyan2011}, which reported low-density regions at \mbox{\feh\ $=-0.2$ dex} and \mbox{\alphafe\ $\sim0.17$ dex}. Although for metallicity, there is an unmistakable minimum around \mbox{\feh\ $=-0.2$ dex}, the same cannot be said for the \sfe\ distribution. The separation would be expected at a \sfe\ ratio close to 0.17 dex, and despite the presence of a smaller number of stars in this bin (compared to the adjacent ones), the remaining distribution is too irregular and does not allow for a confirmation of this separation. 

We performed two-sample Kolmogorov-Smirnov tests on the \sfe\ ratios of the different population samples (thin/thick, thin/\hamr, thick/\hamr) and all of them yielded p-values of magnitude $10^{-8}$ or smaller, rejecting the hypothesis that the stars were drawn from the same population. 

We also obtained stellar ages for our sample from \citet{dmena2019}, which the authors derived using the PARAM v1.3 tool with Parsec isochrones \citep{bressan2012} and \textit{Gaia} DR2 parallaxes (we refer the reader to these papers for more information on the derivation process and for an analysis of abundance-age ratios of various elements). The relationship of \sfe\ ratios and age is shown in the upper panel of Fig. \ref{sfe:age:pops} and the lower panel depicts the \sfe\ vs \feh\ plane with colour according to stellar age. Both panels include only stars with an error in age lower than 1.5 Gyr. The trend of \sfe-age is very similar to other $\alpha$-elements, with a linear increase with age for the thin disk stars and a more exponential increase for the old thick disk stars \citep[see for comparison Fig. 5 in][]{dmena2019}. The \hamr\ stars present intermediate ages between both populations and follow the increase with age of thin disk stars. The thin disk population can be fitted by a slope of $0.011\pm0.001$, which is similar to those given in \citet{dmena2019} for the $\alpha$-elements Mg, Si, Ca, TiI, and TiII. 
% The analysis of Fig. \ref{sfe:age:pops} reveals that older stars belong to the thick disk and are more sulfur-enhanced, whereas younger stars are members of the thin disk and have lower \sfe\ abundance ratios. The \hamr\ stars lie in the transition between these two populations, spanning ages from $\sim$5 to $\sim$11 Gyr. This is the expected distribution for an $\alpha$-element given our current knowledge of the GCE.

% *************
\subsection{Theoretical predictions for [S/Fe]}

% --------------------------- FIGURE ------------------------
\begin{figure}
    \centering
    \includegraphics[width=0.95\hsize]{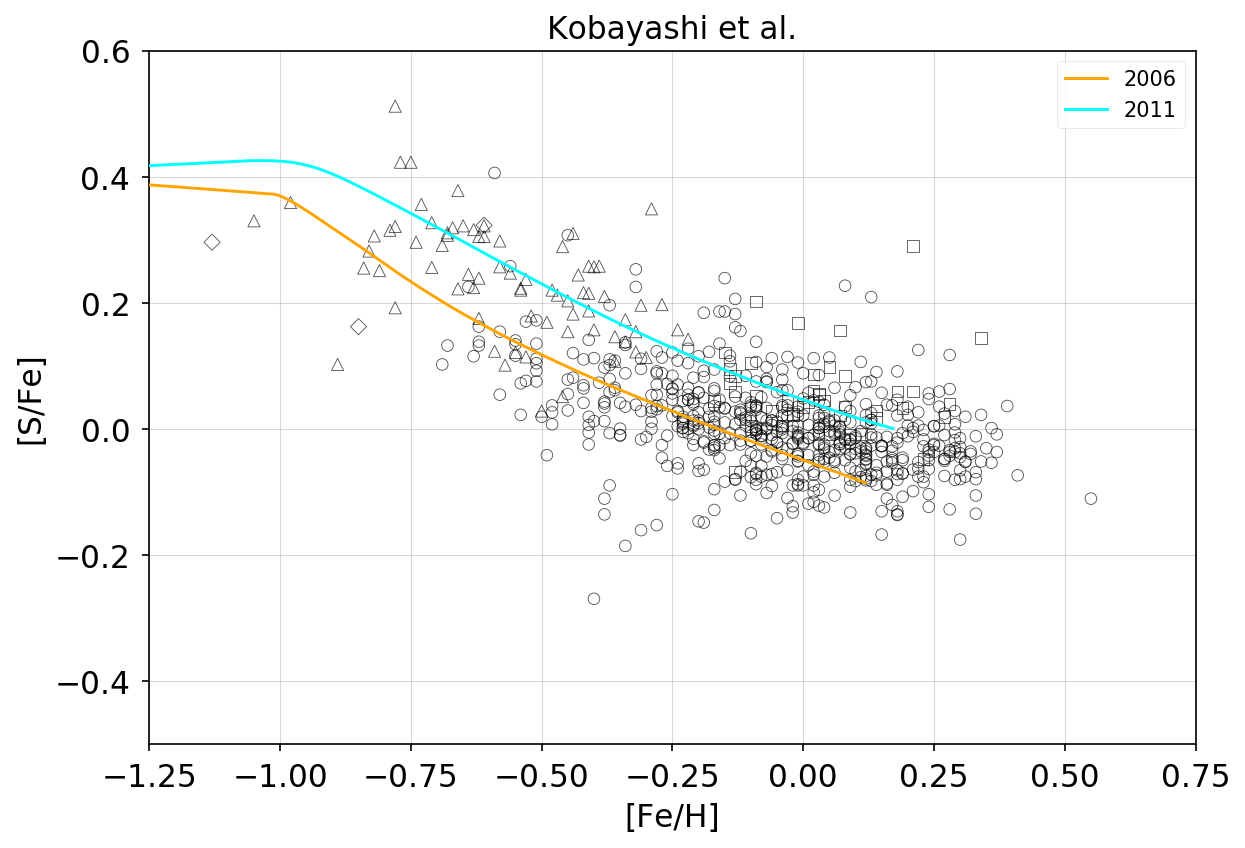}
    \includegraphics[width=0.95\hsize]{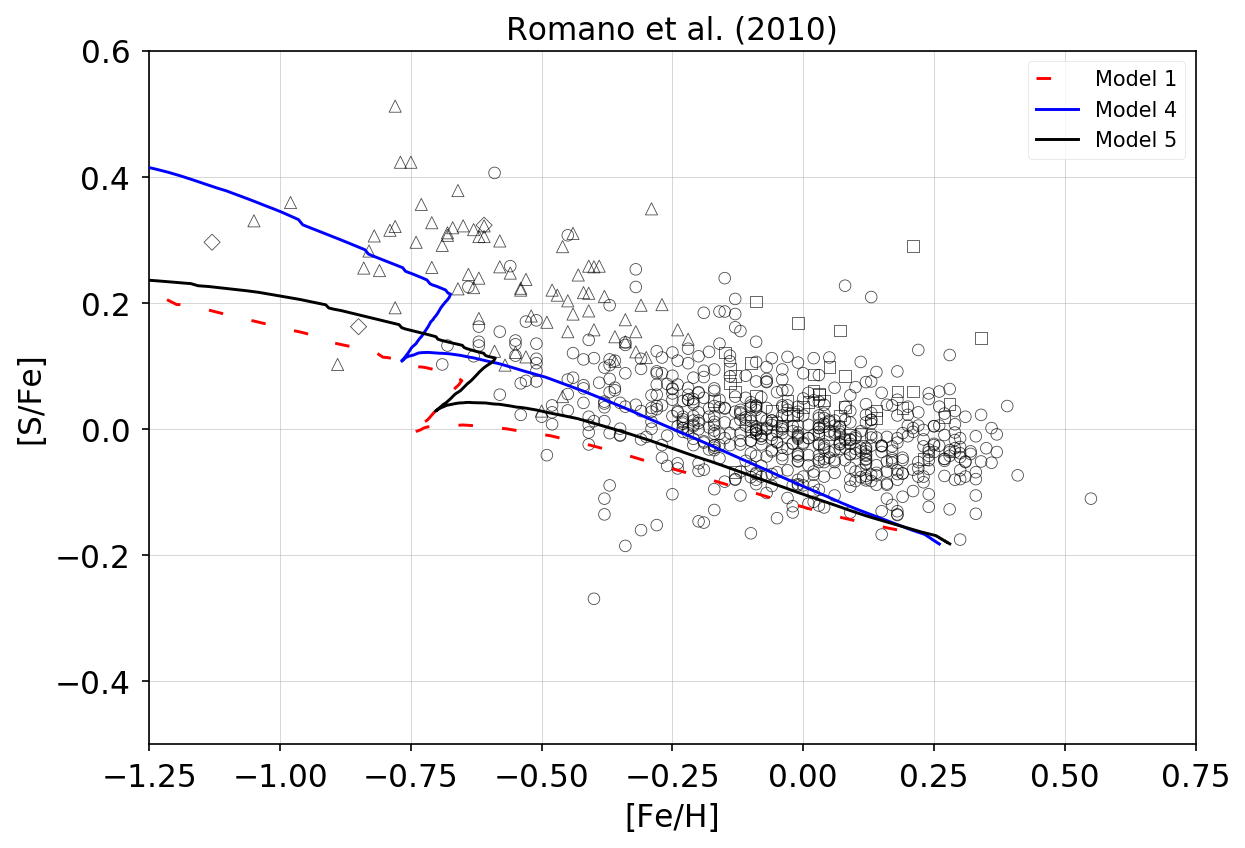}
    \includegraphics[width=0.95\hsize]{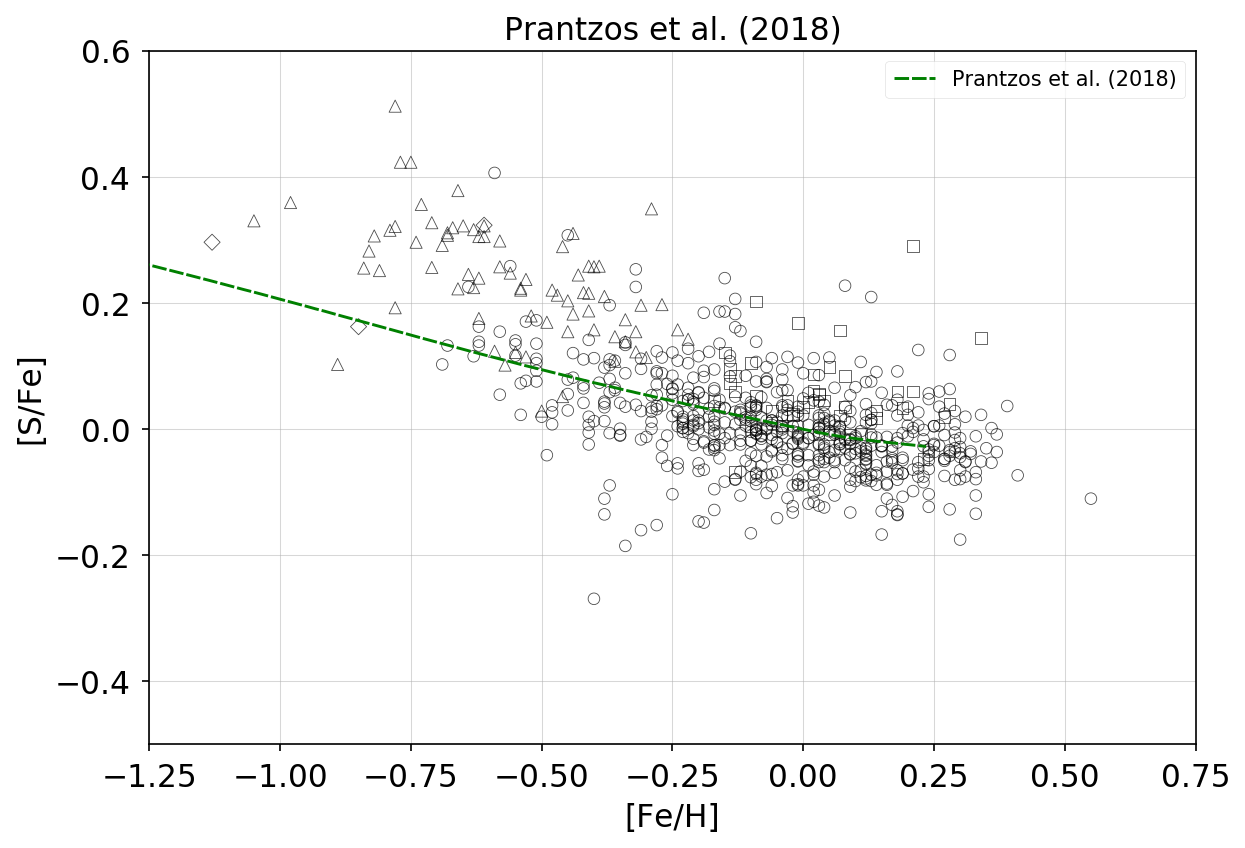}
    \caption{Comparison of observational data to theoretical models. \textit{Top panel:} Models from \citet{kobayashi2006, kobayashi2011}; \textit{Middle panel:} Models 1, 4, and 5 (as numbered in paper) from \citet{romano2010}; \textit{Bottom panel:} Model from \citet{prantzos2018}}
    \label{models}
\end{figure}
% -----------------------------------------------------------

We compare the \sfe\ ratios obtained in this work to model predictions from different authors \citep{kobayashi2006, kobayashi2011, romano2010, prantzos2018} in Fig. \ref{models}. There is a clear distinction in the shape of the lines that represent the evolution of sulfur predicted by the different models. 

% model 1 - yields of Woosley & Weaver (1995) case B
% model 4 - yields of Kobayashi et al. (2006), epsilon=1 for massive stars
% model 5 - yields of Kobayashi et al. (2006), epsilon=0 for massive stars
% prantzos model - yields of Cristallo et al. (2015a) for low and intermediate mass stars; yields of Limongi \& Chieffi (2018) for massive stars, which include mass loss and rotation
% kobayashi 06 - own yields, SNe and HNe, epsilon = 0.5 regardless of mass or metallicity
% kobayashi 11 - updated own yields, AGB, SNe and HNe, epsilon=0.5 for massive stars
% kobayashi private communication - details unknown

The top plot of Fig. \ref{models} shows the models of \citet[hereafter K06]{kobayashi2006} and \citet[hereafter K11]{kobayashi2011}. Both these models consider the yields by SNe and hypernovae (HNe) explosions, but whereas K06 defines the HNe fraction as 0.5 regardless of mass or metallicity of the star, K11 defines a fraction of 0.5 only for stars with mass greater than 20 M$_{\odot}$. K11 also includes yields of asymptotic giant branch (AGB) stars. These two models are the ones that best describe the overall chemical evolution of sulfur. They trace the lower \sfe\ ratios of the thin disk and then raise towards lower metallicities to accompany the thick disk abundances, showing a "knee" feature at \feh\ $\sim-1$, where they go on to form a  plateau. These models are the best fit for the thick disk stars as they accurately fit the \sfe\ ratios at low-metallicity, especially the model from K11.

The middle plot presents three models from \citet{romano2010}. Model 1 (models are numbered as in the paper) adopts the yields of \citet{woosleyweaver1995} case B, while models 4 and 5 adopt the yields of K06, with HNe fractions for massive stars of 1 and 0, respectively. These models underestimate the abundance ratios throughout the entire metallicity range, inaccurately representing the evolution of sulfur. Model 4, however, does present higher \sfe\ values in the very metal-poor regime, which is most likely a product of the high HNe fraction.

Lastly, in the bottom plot, the model from \citet{prantzos2018} is depicted, which adopts the yields of \citet{cristallo2015a} for low and intermediate mass stars and those of \citet{limongichieffi2018} for massive stars, which include mass loss and rotation. This model describes well the \sfe\ ratios of the thin disk, but underestimates the thick disk abundances. A more pronounced slope starting at \feh\ $\leqslant-0.5$ would be expected, following the rise of abundance ratios. This may be due to the fact that no high-energy supernovae explosions (or HNe) were included in the model.

From comparison with our results, the best explanation for the \sfe\ evolution of our sample is a combination of yields from SNe and HNe at high stellar masses and AGB (models by K11).

% *******************
\subsection{Planet hosts vs. non-planet hosts}

% --------------------------- FIGURE ------------------------
\begin{figure}
    \centering
    \includegraphics[width=0.95\hsize]{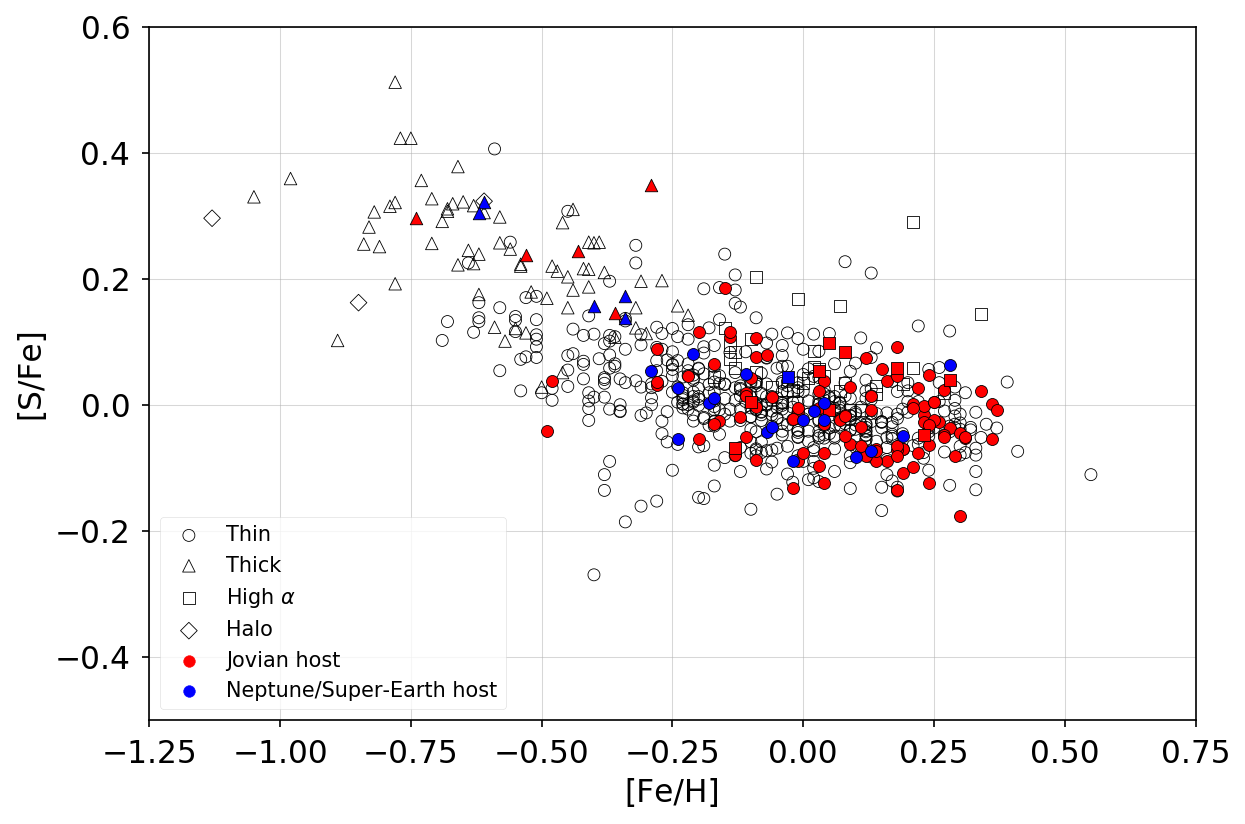}
    \includegraphics[width=0.95\hsize]{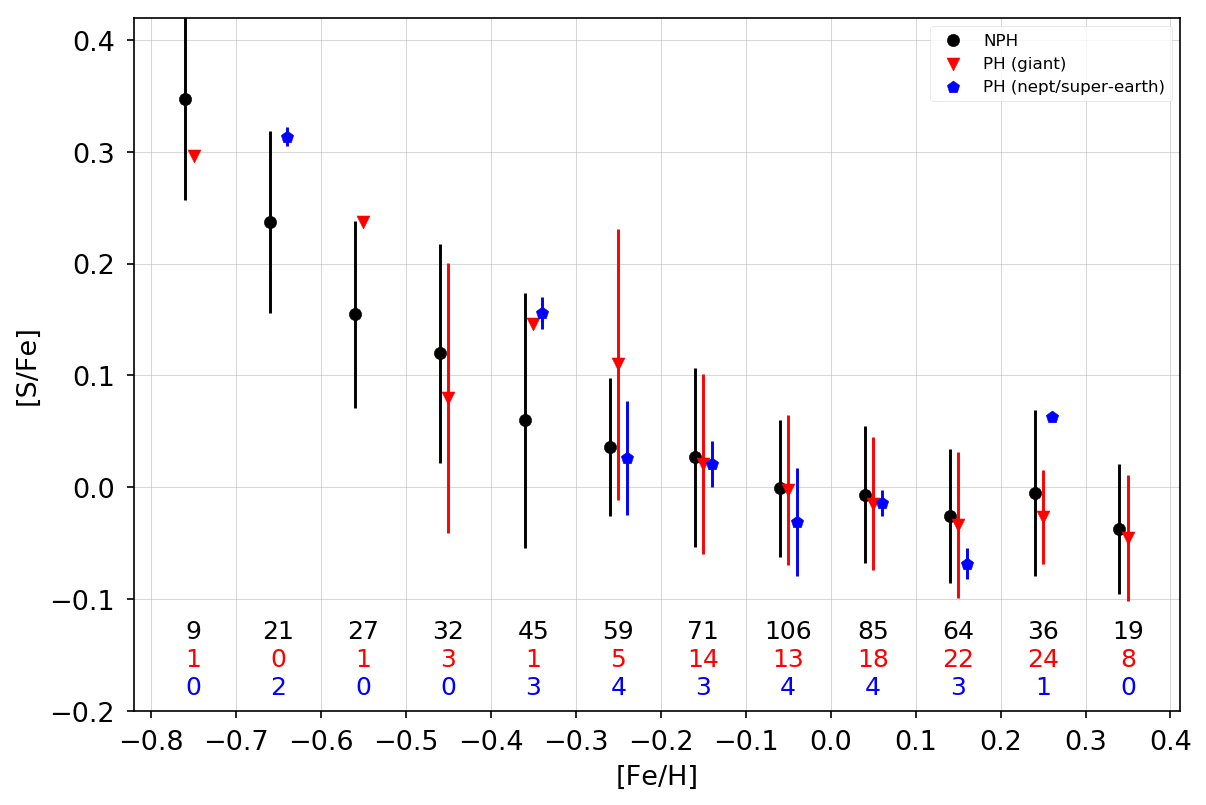}
    \caption{\textit{Top:} Planet hosts vs non-planet hosts, with different colours for planet mass and different symbols for population, as indicated in the legend.
    \textit{Bottom:} Averages of [S/Fe] in each metallicity bin for stars without planets (black circles), jovian-mass hosts (red downwards triangles), and neptunes/super-Earth hosts (blue pentagons).
    }
    \label{sfe:planets}
\end{figure}
% -----------------------------------------------------------

% --------------------------- FIGURE ------------------------
\begin{figure}
    \centering
    \includegraphics[width=0.98\hsize]{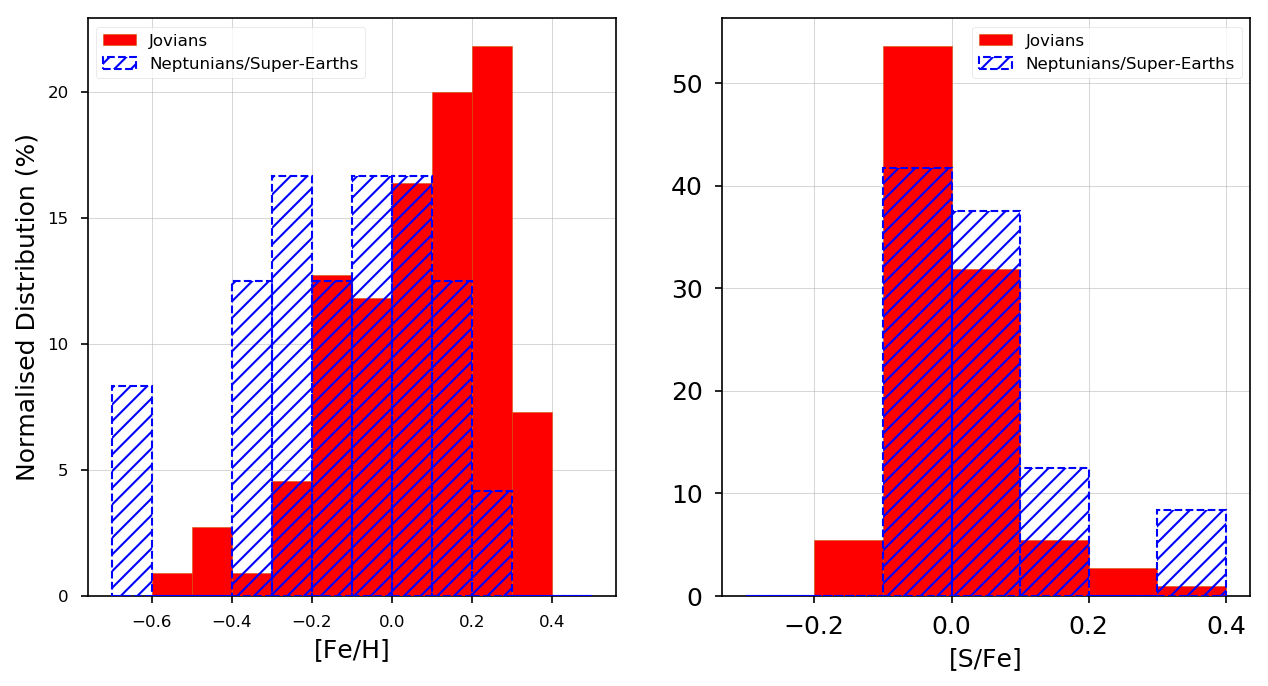}
    \caption{Percentage of stars with giant/neptunian planets within the planet host sample against metallicity (left) and [S/Fe] (right).}
    \label{hist:feh:planets}
\end{figure}
% -----------------------------------------------------------

Our sample of stars with derived sulfur abundances contains 134 stars with known planetary companions, with 110 of them hosting jovian-mass planets while the remaining 24 host Neptune-like planets or Super-Earths. In the top panel of Fig. \ref{sfe:planets}, we can see the distribution of stars with exoplanets (coloured in blue and red) and the comparison sample (empty shapes). The lower panel of Fig. \ref{sfe:planets} displays the average \sfe\ ratios of each of the subsamples (non-planet hosts, giant planet host, and Neptune/Super-Earth planet host) in each metallicity bin. 

The behaviour of the planet hosts (PH) and non-planet hosts (NPH) seems rather similar, not showing any distinctive trends or differences. The \sfe\ ratio averages do not reveal any specific trends of under or overabundance with respect to each other, but rather an irregular distribution for \mbox{\feh$>-0.3$}. 

We note that for \mbox{\feh$<-0.3$}, our average ratios of both types of PH suffer from small-number statistics, and having one single planet host star per bin can, in no way, be taken as representative of the reality. We may be tempted to point out that PH seem to have higher sulfur abundances at lower metallicities, and in fact, with higher number statistics, it has been found that other $\alpha$-elements are indeed enhanced in metal-poor PH (Mg, Si, and Ti, see A12). However, in this work, we need to keep in mind the extremely reduced sample size for this metallicity range. 

Many studies of PH vs NPH show trends of over or under-abundances for other elements, but it is important to note that many of these were based on small samples or rely on comparison samples of 'single' stars that had not been searched for planets, hence not knowing if the stars hosted at least massive planets. Our sample is large enough to allow us to take meaningful conclusions, at least for the range \mbox{\feh$>-0.3$}, regarding the differences (or lack thereof) when comparing sulfur abundances of stars with and without known planetary companions. 

\cite{ecuvillon2004a, ecuvillon2004b} studied volatile elements N, C, S, and Zn in a sample with a metallicity range of $-0.8\leq$ \feh\ $\leq 0.5$ dex and reported no differences in any of these ratios. On the other hand, in \cite{dmena2018}, [Zn/Fe] was found to be overabundant in planet hosts at low metallicity, which is probably driven by the fact that Zn follows the same evolution as $\alpha$-elements, despite not being such an important element for planet-formation as Mg or Si.

Regarding the likelihood of planet formation around certain types of stars, it has been well established that giant planetary companions are much more likely to orbit metal-rich stars, whereas lower mass planets tend to orbit stars of slightly lower metallicity \citep[e.g.][]{sousa2011, buchhave2012}. This tendency can be confirmed in the left-hand plot of Fig. \ref{hist:feh:planets} (only including stars for which we could derive sulfur abundances). On the right-hand plot of the same figure, we find the distribution of planet hosts according to the stars’ \sfe\ ratio. The most evident characteristic is that both giant and Neptune/Super-Earth-sized planets seem to form more often around solar-abundance stars ($-0.1<$\sfe$<0.1$). Additionally, we find that lower mass planets are more likely to orbit stars with higher sulfur ratios (thick disk) rather than lower (thin disk).

%%%%%%%%%%%%%%%%%%%%%%%%%%%%%%%%%%%%%%%%%%%%%%%%%%%%%%%%%%%%%%%%%%%%%%%%
\section{Summary}

We performed spectral synthesis on a sample of 1059 \mbox{HARPS-GTO} solar neighbourhood stars, for lines of Multiplet 8 (triplets at \mbox{6743 $\AA$} and \mbox{6757 $\AA$}), having obtained \sfe\ ratios for 719 of the stars and for the Sun, through a Vesta-reflected spectrum. A third triplet at \mbox{6748 $\AA$} was considered, but discarded due to incorrect fitting of the models. Values of rotational velocity were also determined for most stars of the initial sample. 

The results were interpreted in the context of the \sfe\ vs \feh\ plane, the distribution of sulfur in Galactic populations, how sulfur abundances relate to stellar age, the evolution of sulfur according to theoretical models of Galactic chemical evolution and, finally, comparing abundances of planet-hosts and non-planet hosts. Our conclusions can be summarised as follows:

\begin{enumerate}
    \item In the \sfe\ vs \feh\ plane, sulfur appears to behave like a typical $\alpha$-element, showing the expected trend  with solar-abundances around solar metallicity and a rise towards lower metallicities. There is an apparent decrease of the \sfe\ ratios towards higher metallicities, although this should be taken with caution due to having few stars at this extreme. Similarly, we can not confidently state that the ratios form a plateau for \mbox{\feh$<-1.0$ dex} due to the small number of stars below this threshold.
    \item The distribution of sulfur ratios in the Galactic populations is also $\alpha$-like. The thin and thick disks overlap slightly in the chemical plane but are clearly separated overall. Sulfur-enhanced and older stars are identified as belonging to the thick disk population whereas younger stars are sulfur-poor and categorised as thin disk members. 
    \item The $\alpha$ separation line \citep[presented in][and updated in DM17]{adibekyan2011} was not able to visually separate the low- and high-$\alpha$ sequence in a very accurate manner, but this can be due to large uncertainties in sulfur abundances and differences in the nucleosynthesis of this element compared to those of other $\alpha$-elements. Nevertheless, it has been confirmed that the high-$\alpha$ metal-rich stars are, on average, sulfur-enhanced compared to the thin disk counterparts of the same metallicity, similarly to their enhancement of other $\alpha$-elements. Statistically, the Kolmogorov-Smirnov tests performed on the \sfe\ ratios of the thin disk, thick disk, and \hamr\ group reject the hypothesis that the stars were drawn from the same population.
    \item The \sfe-age relation is very clear and similar to those found for other $\alpha$-elements in \cite{dmena2019}, reinforcing that $\alpha$-abundances are a good proxy for stellar age.
    \item In the metal-poor regime, the \sfe\ vs \feh\ distribution is best fitted by the models of \cite{kobayashi2006, kobayashi2011}, as they follow the rise of the [S/Fe] ratios towards lower metallicities, placing the ‘knee’ of the thick disk evolution at around \feh$=-1.0$. These models included normal energy and high energy supernovae explosions, although in different fractions, and \citet{kobayashi2011} also included AGB yields. As for the metal-rich tail, the evolution of the sulfur ratios are best described by the model of \cite{prantzos2018}, who considered only "typical" energy supernovae. Nonetheless, we note that none of the models reach the high metallicities presented in this work.
    \item We find no evidence of under or overabundance of \sfe\ when comparing the average ratios of stars with and without planetary companions. This lack of differences between the two groups of stars was also reported in \cite{ecuvillon2004a, ecuvillon2004b} for other volatile elements. For metallicities below -0.3 dex, firm conclusions regarding the abundance comparisons cannot be reached due to the small number of planet host stars in these metallicity bins. Even so, for the few metal-poor planet hosts for which we could derive \sfe\ ratios, it is enhanced, despite the difference with respect to thick disk stars without planetary companions not being relevant.
    \item Planet companions are more common around stars with solar-abundances of sulfur, regardless of the planet mass. Lower mass planets seem more likely to orbit sulfur-enhanced stars rather than sulfur-depleted, although still with much lower frequency than around solar-abundance stars.
    
\end{enumerate}

All in all, sulfur can be considered as a typical $\alpha$-element for all purposes of Galactic chemical evolution, Galactic populations and ages. With regard to planet host stars, abundances of this volatile element are no more abundant or scarce than in those without planetary companions. There is a necessity for future studies covering the low-metallicity range, which will help clarify some details which could not be analysed in our data, as well as place further constrains in important formation mechanisms of our Universe. Moreover, it would be interesting to have models of \sfe\ evolution for \feh\ $\geqslant 0.25$ dex, since our data seem to point to a continued decrease at high metallicities, but the models which best fit our sample in this region \citep{prantzos2018} present a rather flattened trend with increasing metallicity.

%%%%%%%%%%%%%%%%%%%%%%%%%%%%%%%%%%%%%%%%%%%%%%%%%%%%%%%%%%%%%%%%%%%%%%%%
\begin{acknowledgements}
     We thank Alejandra Recio-Blanco and Patrick de Laverny for fruitful discussion, Vardan Adibekyan for useful suggestions, and the anonymous referee for helpful comments which helped improve this paper. A.R.C.S., E.D.M. and M.T. acknowledge the support by FCT/MCTES through national funds (PIDDAC) by this grant UID/FIS/04434/2019. This work was also supported by FCT - Fundação para a Ciência e a Tecnologia through national funds (PTDC/FIS-AST/7073/2014) and by FEDER - Fundo Europeu de Desenvolvimento Regional through COMPETE2020 - Programa Operacional Competitividade e Internacionalização (POCI--01--0145-FEDER--016880). A.R.C.S. acknowledges the support by the Student Initiation Fellowship funded by the grant UID/FIS/04434/2013 \& POCI--01--0145-FEDER--007672. E.D.M. acknowledges the support by the Investigador FCT contract IF/00849/2015/CP1273/CT0003 and in the form of an exploratory project with the same reference. 
\end{acknowledgements}

%------------------------------------------------------------
\bibliographystyle{aa}
\bibliography{sulfur.bib}

\begin{appendix}

\section{Comparison to literature}

\begin{table*}[]
% \begin{threeparttable}
\centering
\caption{Comparison of [S/Fe] ratios derived in this work with values from the literature.}
\label{tab:my-table}
\begin{tabular}{lccc}
\hline
 & \sfe$_{literature}$ & [S/Fe]$_{this\ work}$ & Reference \\ \hline
 
HD4307 & 0.03 $\pm$ 0.11 & 0.01 $\pm$ 0.03 &  [1] \\
       & -0.08 $\pm$ 0.10 & 0.01 $\pm$ 0.03 &  [2] \\
HD10700 & 0.06 $\pm$ 0.11 & 0.18 $\pm$ 0.14 &  [1] \\
HD16141 & -0.04 $\pm$ 0.10 & -0.09 $\pm$ 0.02 &  [3] \\
        & -0.08 $\pm$ 0.11 & -0.09 $\pm$ 0.02 &  [1] \\
HD16548 & -0.12 $\pm$ 0.02 & -0.17 $\pm$ 0.03 & [3] \\
HD19994 & -0.10 $\pm$ 0.11 & -0.03 $\pm$ 0.04 & [1] \\
HD21019 & 0.10 $\pm$ 0.03 & 0.03 $\pm$ 0.05 & [3] \\
        & -0.01 $\pm$ 0.11 & 0.03 $\pm$ 0.05 & [1] \\
HD22049 & 0.02 $\pm$ 0.11 & 0.19 $\pm$ 0.09 &  [1] \\
        & 0.04 $\pm$ 0.10 & 0.19 $\pm$ 0.09 & [2] \\
HD22879 & 0.14 $\pm$ 0.10 & 0.31 $\pm$ 0.08 &  [4] \\
HD23249 & 0.24 $\pm$ 0.11 & 0.16 $\pm$ 0.08 &  [3] \\
HD24892 & 0.19 $\pm$ 0.03 & 0.12 $\pm$ 0.05 &  [3] \\
HD28185 & -0.07 $\pm$ 0.09 & 0.00 $\pm$ 0.03 & [3] \\
HD38858 & -0.11 $\pm$ 0.10 & 0.05 $\pm$ 0.04 & [2] \\
HD44420 & 0.12 $\pm$ 0.18 & -0.01 $\pm$ 0.03 &  [5] \\
HD49035 & -0.05 $\pm$ 0.10 & -0.04 $\pm$ 0.03 & [4] \\
HD52265 & -0.05 $\pm$ 0.11 & -0.01 $\pm$ 0.03 &  [1] \\
HD59984 & 0.04 $\pm$ 0.11 & 0.10 $\pm$ 0.04 &  [1] \\
HD69830 & -0.09 $\pm$ 0.11 & -0.04 $\pm$ 0.06 & [1] \\
HD76151 & 0.09 $\pm$ 0.11 & -0.04 $\pm$ 0.03 &  [1] \\
        & 0.04 $\pm$ 0.10 & -0.04 $\pm$ 0.03 & [2] \\
HD79601 & 0.01 $\pm$ 0.10 & 0.12 $\pm$ 0.05 &  [4] \\
HD82943 & 0.04 $\pm$ 0.18 & -0.03 $\pm$ 0.02 &  [5] \\
        & -0.07 $\pm$ 0.11 & -0.03 $\pm$ 0.02 & [1] \\
HD90722 & 0.03 $\pm$ 0.12 & 0.02 $\pm$ 0.04 &  [5] \\
HD92788 & 0.07 $\pm$ 0.17 & -0.05 $\pm$ 0.03 & [5] \\
HD106116 & 0.05 $\pm$ 0.16 & -0.06 $\pm$ 0.03 & [5] \\ 
HD107148 & -0.06 $\pm$ 0.06 & -0.05 $\pm$ 0.03 &  [3] \\
HD114613 & 0.01 $\pm$ 0.04 & -0.07 $\pm$ 0.02 &  [3] \\
HD115617 & -0.14 $\pm$ 0.11 & -0.09 $\pm$ 0.04 & [1] \\
HD125184 & -0.07 $\pm$ 0.11 & -0.08 $\pm$ 0.04 &  [1] \\
         & -0.13 $\pm$ 0.10 & -0.08 $\pm$ 0.04 &  [2] \\
HD134987 & -0.04 $\pm$ 0.11 & 0.00 $\pm$ 0.04 &  [1] \\
HD144585 & -0.06 $\pm$ 0.11 & -0.14 $\pm$ 0.03 &  [1] \\
HD146233 & -0.04 $\pm$ 0.10 & -0.05 $\pm$ 0.03 & [2] \\
HD161098 & -0.09 $\pm$ 0.10 & -0.03 $\pm$ 0.05 &  [2] \\
HD169830 & -0.09 $\pm$ 0.11 & -0.08 $\pm$ 0.03 & [1] \\
HD179949 & -0.10 $\pm$ 0.11 & -0.10 $\pm$ 0.03 & [1] \\
HD199960 & -0.04 $\pm$ 0.11 & -0.04 $\pm$ 0.03 &  [1] \\
         & -0.04 $\pm$ 0.10 & -0.04 $\pm$ 0.03 &  [2] \\
HD207970 & -0.04 $\pm$ 0.10 & 0.02 $\pm$ 0.04 &  [4] \\
HD210277 & -0.01 $\pm$ 0.06 & 0.06 $\pm$ 0.04 & [3] \\
         & -0.06 $\pm$ 0.11 & 0.06 $\pm$ 0.04 &  [1] \\ 
HD211038 & 0.44 $\pm$ 0.04 & 0.20 $\pm$ 0.08 & [3] \\
HD219077 & 0.21 $\pm$ 0.03 & -0.07 $\pm$ 0.05 &  [3] \\
HD221420 & -0.08 $\pm$ 0.05 & -0.11 $\pm$ 0.03 &  [3] \\
\\

\end{tabular}

\tablefoot{[1] \cite{takeda2016}; [2] \cite{mishenina2016} (errors in abundance ratios taken from average error bars in corresponding plot of their Fig. 5); [3] \cite{maldonado2016}; [4] \cite{duffau2017}; [5] \cite{caffau2019}.} 

\end{table*}

\end{appendix}

\end{document}